\documentclass{aa}  

\usepackage{amsmath}
\usepackage{amssymb}
\usepackage{graphicx}
\usepackage{color}
\usepackage{txfonts}
\usepackage{comment}
\usepackage{ulem}

\usepackage{xcolor}
\definecolor{xlinkcolor}{cmyk}{1,1,0,0}
 \usepackage[bookmarks=false,         
     pdfnewwindow=true,      
     colorlinks=true,    
     linkcolor=xlinkcolor,     
     citecolor=xlinkcolor,     
     filecolor=xlinkcolor,  
     urlcolor=xlinkcolor,      
final=true
 ]{hyperref}

\usepackage{savesym}
\savesymbol{tablenum}
\usepackage{siunitx}
\usepackage{xspace}
\restoresymbol{SIX}{tablenum}
\usepackage{hyperref}

\newcommand\dbquote[1]{\textquotedblleft #1\textquotedblright}

\newcommand\partialdiff[1]{\frac{\partial}{\partial #1}}

\newcommand\partiallogdiff[1]{\textrm{d}\mathrm{ln} #1/\textrm{d} \mathrm{ln}\, r}

\AtBeginDocument{}%
\AtBeginDocument{}%
\AtBeginDocument{}%
\AtBeginDocument{}%
\AtBeginDocument{}%

\begin{document} 
   \title{Large gaps and high accretion rates in photoevaporative transition disks with a dead zone}
   \author{Mat\'ias G\'arate\inst{1, 2}
          \and Timmy N. Delage\inst{1}
          \and Jochen Stadler\inst{1}
          \and Paola Pinilla\inst{1,3}
          \and Til Birnstiel\inst{2,4}
          \and Sebastian Markus Stammler\inst{2}
          \and Giovanni Picogna\inst{2}
          \and Barbara Ercolano\inst{2}
          \and Raphael Franz\inst{2}
          \and Christian Lenz\inst{1}
          }
   \institute{
   $^{1}$Max-Planck-Institut f\"ur Astronomie, K\"onigstuhl 17, 69117, Heidelberg, Germany\\
   $^{2}$University Observatory, Faculty of Physics, Ludwig-Maximilians-Universit\"at M\"unchen, Scheinerstr.\ 1, 81679 Munich, Germany\\
   $^{3}$ Mullard Space Science Laboratory, University College London, Holmbury St Mary, Dorking, Surrey RH5 6NT, UK\\
   $^{4}$Exzellenzcluster ORIGINS, Boltzmannstr. 2, D-85748 Garching, Germany\\
              \email{garate@mpia.de}
             }
   \date{}

  \abstract
  {Observations of young stars hosting transition disks show that several of them have high accretion rates, despite their disks presenting extended cavities in their dust component. This represents a challenge for theoretical models, which struggle to reproduce both features simultaneously.}
  {We explore if a disk evolution model, including a dead zone and disk dispersal by X-ray photoevaporation,  can explain the high accretion rates and large gaps (or cavities) measured in transition disks.}
  %
  {We implement a dead zone turbulence profile and a photoevaporative mass loss profile into numerical simulations of gas and dust. We perform a population synthesis study of the gas component, and obtain synthetic images and SEDs of the dust component through radiative transfer calculations.}
  {This model results in long lived inner disks and fast dispersing outer disks, that can reproduce both the accretion rates and gap sizes observed in transition disks.
  For a dead zone of turbulence $\alpha_\textrm{dz} = \SI{e-4}{}$ and an extent $r_\textrm{dz} = \SI{10}{AU}$, our population synthesis study shows that $63\%$ of our transition disks are still accreting with $\dot{M}_\textrm{g} \geq \SI{e-11}{M_\odot\, yr^{-1}}$ after opening a gap.
  Among those accreting transition disks, half display accretion rates higher than $\SI{5.e-10}{M_\odot\, yr^{-1}}$.
  The dust component in these disks is distributed in two regions: in a compact inner disk inside the dead zone, and in a ring at the outer edge of the photoevaporative gap, which can be located between $\SI{20}{AU}$ and $\SI{100}{AU}$. 
  Our radiative transfer calculations show that the disk displays an inner disk and an outer ring in the millimeter continuum, a feature that resembles some of the observed transition disks.}
  {A disk model considering X-ray photoevaporative dispersal in combination with dead zones can explain several of the observed properties in transition disks including: the high accretion rates, the large gaps, and a long-lived inner disk at mm-emission. }


   \maketitle
%

\section{Introduction} \label{sec_Intro}
The nature of the observed structures in protoplanetary disks and their relation to the disk evolution has been a subject of study for over 30 years now.
Transition disks, in particular, are disks that present a deficit in the near-infrared (NIR) and/or mid-infrared (MIR) emission, while still displaying the characteristic far-infrared (FIR) excess of most protoplanetary disks \citep[][]{Strom1989, Skrutskie1990}, and represent one of the key puzzle pieces to understand the disk evolution process \citep[see reviews by][]{Owen2016_review, Ercolano2017_Review}.\par
The lack of NIR/MIR emission in transition disks' spectral energy distribution (SED) is attributed to a gap in the dust component, or more precisely, to the lack of hot micron sized grains in the inner disk \citep[see][for a review]{Espaillat2014}. 
Through radiative transfer models it has been possible to measure the radial extent of the dust cavity for a wide sample of transition disks \citep[][]{vanderMarel2016} and observations at different wavelengths have also shown that some of these objects retain a compact dust component close to the star \citep[e.g.][]{Espaillat2010, Benisty2010, Olofsson2013, Matter2016, Kluska2018, Pinilla2019, Pinilla2021}, demonstrating that for some transition disks the gap is not completely devoid of dust.\par 
One particularly curious and challenging feature of transition disks is that a large number of them show gas accretion signatures, with rates that can be as high as $\dot{M}_g \sim \SI{e-8}{M_\odot \, yr^{-1}}$ \citep[e.g.][]{Cieza2012, Alcala2014, Manara2014, Manara2017}, indicating that there is plenty of gaseous material close to the star, or that low density material accretes at supersonic speeds into the central star, driven by a combination of winds and magneto-hydrodynamic (MHD) processes \citep[e.g.][]{Wang2017}. 
If the inner cavities are really rich in gas, then this raises the following question: How is it possible to create a prominent dust gap dust, while still retaining a long lived inner gas disk that can produce such accretion rates?\par
Classical theoretical models suggest that disk evolution occurs in two different stages: A viscous evolution stage, where the transport of angular momentum drives the accretion of gas onto the star \citep[][]{Lynden-Bell1974, Pringle1981}, and a dispersal stage, where the mass loss rate due to photoevaporation overcomes the accretion rate onto the star, creating a cavity and dispersing the disk from the inside out \citep[][]{Clarke2001, Alexander2006a, Alexander2006b, Alexander2007}.\par
Though photoevaporation models can easily predict large gaps that extend for tens of AU, these fail to explain the observed high accretion rates, since the gaseous inner disks are short-lived and quickly accreted once the gap opens \citep[][]{Owen2010, Owen2011, Picogna2019}. As a consequence, these models tend to over-predict the fraction of non-accreting disks with gaps that extend beyond $\SI{20}{AU}$, although the mechanisms that speed up the depletion of the outer disk, such as thermal sweeping and depletion of carbon and oxygen, can partially alleviate the discrepancy with observations \citep[][]{Owen2013, Ercolano2018, Wolfer2019}.\par
On the other hand, planet-disk interactions allow for high accretion rates in the inner disk \citep{Lubow2006}, and also create a gap where the large dust particles are trapped at its outer edge \citep[][]{Pinilla2012}. Depending on the planet mass and disk properties (e.g. disk viscosity), the inner disk can be replenished of micron-sized particles from the outer disk. However, up to today planets have been only detected in the cavity of one transition disk \citep[PDS 70,][]{Keppler2018, Christiaens2019}, while current observational capabilities should already have detected some of the planets inferred to explain the structures of some of them \citep[e.g.][]{AsensioTorres2021}. This questions the universality of planets as the potential origin of the transition disks like structures. 
Invoking multiple planets can explain a wider variety of structures \citep[][]{Pinilla2015}, but it can also reduce the lifetime of the inner disk, leading to the same problems as photoevaporation models \citep[][]{Zhu2011}.\par
Other models including grain growth and/or dead zones \citep[i.e. regions with low turbulent viscosity,][]{Gammie1996}, show that these can create transition disk structures \citep{Dullemond2005, Birnstiel2012_b, Flock2015} comparable to the ones observed at different wavelengths \citep[e.g.][]{Regaly2012, Pinilla2016}.\par
In this work we aim to explain the accretion rates and gap sizes observed in transition disks by revisiting and expanding the model of \cite{Morishima2012}, which studied the evolution of disks with a dead zone undergoing photoevaporative dispersal. In their model, a protoplanetary disk is expected to evolve through the following steps:
\begin{itemize}
    \item The disk viscous evolution is driven  by the turbulence created by the magneto rotational instability \citep[MRI,][]{Balbus1998}.
    \item In the inner regions the ionization fraction is low, creating a \dbquote{dead zone} where the MRI is inefficient, and the turbulent viscosity is low \citep[][]{Gammie1996}.
    \item Because of the difference in turbulent viscosity, the inner disk evolves slower than the outer disk. This causes the accretion rate at the outer \dbquote{active} regions to decrease faster than in the inner \dbquote{dead} regions.
    \item Photoevaporation clears a gap  outside the dead zone, once the accretion rate has dropped below the mass loss rate. Meanwhile, the inner disk is unaffected by photoevaporation. 
    \item This results in a long lived inner disk that acts as an accretion reservoir, and a large gap in the outer disk that continues to expand from the inside out due to photoevaporation \citep[][]{Morishima2012, Bae2013}.
\end{itemize}
The previous study of \cite{Morishima2012} demonstrated that large gaps and high accretion rates can be recreated through the combined effects of dead zones and photoevaporation, however it is still unclear if this model can produce a fraction of accreting transition disks that is consistent with observations \citep[see][]{Hardy2015, Owen2016_review}, what is the predicted distribution of accretion rates and cavity sizes, and if the dust component can reproduce the characteristic features of transition disks, such as the deficit in the NIR/MIR emission.\par
Motivated by these open questions, we construct a series of population synthesis models, where we include the state-of-the-art X-ray photoevaporation model from \cite{Picogna2019} and a parametric dead zone prescription, and implement them into the modular disk evolution code \texttt{DustPy} (Stammler and Birnstiel, in prep.). 
For each population we want to measure the gas accretion rate and the gap size distribution, compare it with the observed population of transition disks and determine which dead zone properties are more likely to produce high accretion rates with large gaps, without overestimating the fraction of non-accreting disks.\par
To test if this model can produce a signal that is consistent with observations of transition disks, we also include a dust coagulation model, where we track the evolution of the grain size distribution during disk dispersal, and obtain synthetic SEDs and images at both NIR and millimeter wavelengths.\par
%
This paper is organized as follows. In \autoref{sec_Model} we present our disk evolution model. 
In \autoref{sec_Setup} we describe our numerical setup of our simulations.
In \autoref{sec_Results_Gas} we show the results of our population synthesis study, focusing on the gas evolution.
\autoref{sec_Results_Dust} shows the evolution of a single simulation including the evolution of dust, along with synthetic images from a radiative transfer calculation.
In \autoref{sec_Discussion} we discuss how our model relates to the observed population of transition disks, its caveats and follow up work.
We summarize our results in \autoref{sec_Summary}.
%
%
\section{Disk evolution model} \label{sec_Model}
For our model we consider an axisymmetric protoplanetary disk undergoing viscous accretion, and describe its gas surface density evolution in 1D using the following diffusion equation \citep{Lust1952, Lynden-Bell1974, Pringle1981}:
\begin{equation} \label{eq_GasEvolution}
    \partialdiff{t}\Sigma_\textrm{g} = \frac{3}{r}\partialdiff{r}\left(r^{1/2} \partialdiff{r}\left(\nu \Sigma_\textrm{g} r^{1/2} \right) \right) - \dot{\Sigma}_w,
\end{equation}
where $\Sigma_\textrm{g}$ is the gas surface density, $r$ is the radial distance to the star (in cylindrical coordinates), $\nu$ is the kinematic viscosity and $\dot{\Sigma}_w$ is the mass loss rate due to photoevaporation.\par
For the kinematic viscosity we adopt the \cite{Shakura1973} model:
\begin{equation} \label{eq_nu_alpha}
    \nu = \alpha c_s h_\textrm{g},
\end{equation}
where $\alpha$ is a dimensionless parameter that controls the intensity of the disk turbulence, $h_\textrm{g}$ is the gas scale height, and $c_s = \sqrt{\gamma k_B T / \mu m_p}$ is the sound speed, with $k_B$ the Boltzmann constant, $T$ the disk temperature, $\mu = 2.3$ the mean molecular weight, $m_p$ the proton mass and $\gamma = 1.4$ the adiabatic index.\par
In terms of advection, the radial accretion rate of a gas disk is $\dot{M}_\textrm{g} = 2 \pi r \Sigma_\textrm{g} v_\textrm{g}$, where $v_\textrm{g}$ corresponds to the gas viscous velocity:
\begin{equation} \label{eq_velocity_visc}
v_\textrm{g} = -3 \alpha c_s \frac{h_\textrm{g}}{r}\, \frac{\textrm{d}\mathrm{ln}}{ \textrm{d} \mathrm{ln}\, r}\left(\nu \, \Sigma_\textrm{g} \, \sqrt{r}\right).
\end{equation}\par
Since the viscous evolution is controlled by the $\alpha$ parameter, we can already infer that regions with low turbulence evolve slowly, and regions with high turbulence evolve rapidly.
%
\subsection{Photoevaporation model}
\label{sec_Model_Photo}
\begin{figure}
\centering
\includegraphics[width=90mm]{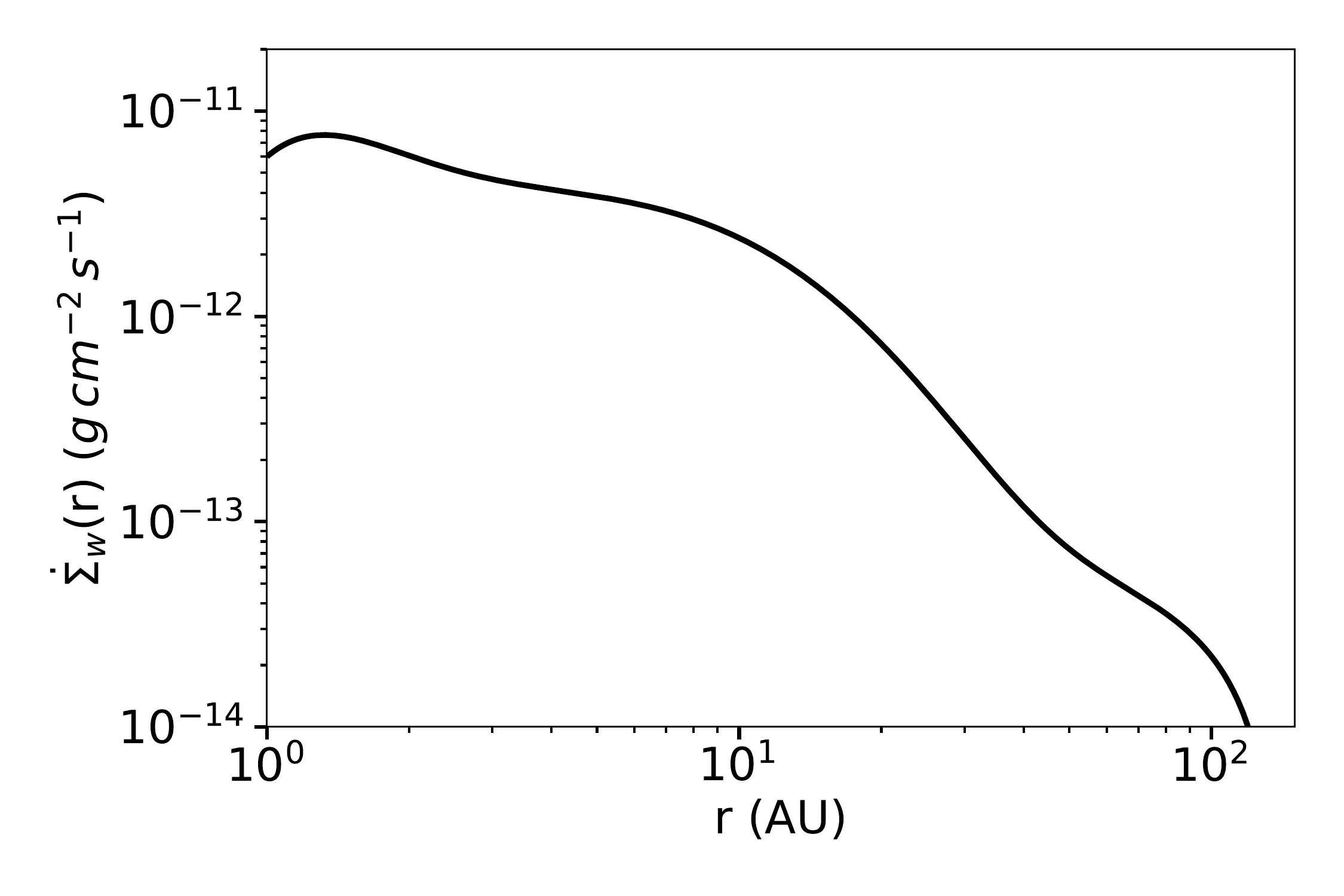}
 \caption{
 Surface density loss rate due to X-ray photoevaporation, for a star with $L_x = \SI{e30}{erg\, s^{-1}}$, following \cite{Picogna2019}.
 }
 \label{Fig_PhotoModel}
\end{figure}

Photoevaporation is believed to play a major role in the dispersal of protoplanetary disks. As the accretion rate decreases over time due to viscous evolution, disks naturally transition to a photoevaporating regime where a gap opens from inside out, once the mass loss rate surpasses the local gas accretion rate \citep{Clarke2001, Alexander2006a, Alexander2006b}.\par
For this study, we focus on the effects of X-ray photoevaporation \citep{Ercolano2008a, Ercolano2009, Owen2010}, and implement the mass loss rate profiles from \cite{Picogna2019} into the sink term $\dot{\Sigma}_w$ of \autoref{eq_GasEvolution}.\par
We refer the reader to the original \citet[][their Eqs. 2-5]{Picogna2019} for a detailed description of the mass loss rate, which is derived from 2D hydrodynamic models with radiative transfer calculations, and parameterized as a function of the stellar X-ray luminosity $L_x$. \autoref{Fig_PhotoModel} shows the corresponding $\dot{\Sigma}_w$ profile used for this work.\par
For reference, in this model an X-ray luminosity of $L_x = \SI{e30}{erg\, s^{-1}}$ corresponds to a total mass loss rate of $\dot{M}_w \approx \SI{e-8}{M_\odot\, yr^{-1}}$, and is overall higher than the previous photoevaporative model from \cite{Owen2010}, making it easier to open a large gap in the gas.
%
%
%
\subsection{Dead zone model} \label{sec_Model_DZ}
\begin{figure}
\centering
\includegraphics[width=90mm]{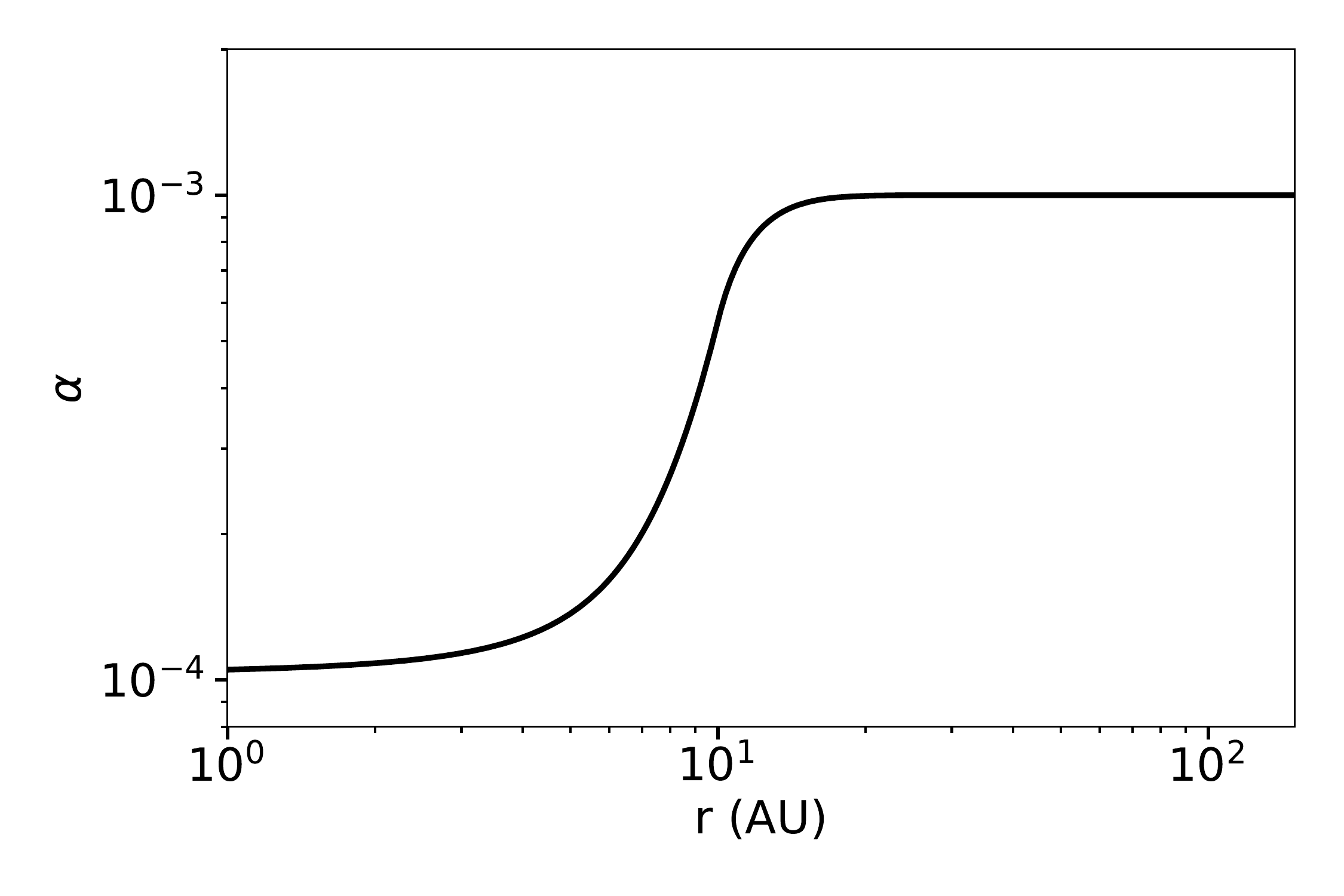}
 \caption{
  $\alpha$ parameter radial profile from \autoref{eq_alpha_profile}, using $\alpha_\textrm{a}= \SI{e-3}{}$, $\alpha_\textrm{dz}= \SI{e-4}{}$, and $r_\textrm{dz}= \SI{10}{AU}$.
 }
 \label{Fig_AlphaModel}
\end{figure}
Turbulence driven by the magneto rotational instability (MRI) is one of the candidates to explain the angular momentum transport across protoplanetary disks \citep[][]{Balbus1998}, and is triggered by the coupling between the charged particles in the gas phase and the magnetic field \citep[][]{Balbus1991}.\par
In the inner regions of the disk, where the column density is higher and the ionization lower, the MRI is quenched or even shut off, creating a \dbquote{dead zone} \citep[][]{Gammie1996}.
In contrast with the MRI active regions, the turbulence in the dead zone is much lower, and is dominated by less efficient hydrodynamic instabilities, such as the vertical shear instability \citep[VSI,][]{Nelson2013, Flock2020, Manger2020}.\par
In our 1D model, we parameterize the turbulence across the protoplanetary disk using the following profile:
\begin{equation}\label{eq_alpha_profile}
\alpha(r) = \alpha_\textrm{dz} + (\alpha_\textrm{a} - \alpha_\textrm{dz}) \cdot \begin{cases}
            \frac{1}{2} \exp\left(\Delta r / w\right)    &    r < r_\textrm{dz}\\
            1 - \frac{1}{2} \exp\left(-\Delta r / w\right)   &    r \geq r_\textrm{dz},
      \end{cases}
\end{equation}
where $\alpha_\textrm{a}$ is the turbulence parameter of the MRI active region, $\alpha_\textrm{dz}$ is the turbulence parameter of the dead zone, $r_\textrm{dz}$ is the dead zone radial extent, $\Delta r = r - r_\textrm{dz}$, and $w = r_\textrm{dz}/5$ is the transition width from between the dead and the active regions; similar to the models used by \cite{Birnstiel2012_b, Garate2019}. For reference, we show the profile in \autoref{Fig_AlphaModel}.\par
This profile results in a fast viscous evolution for the outer disk ($r > r_\textrm{dz}$), and a slow viscous evolution for the dead zone in the inner disk ($r < r_\textrm{dz}$).\par
While this profile does not include the dependence on the surface density profile from other models \citep[e.g.][]{Kretke2009, Morishima2012, Pinilla2016}, it makes it easier to study the impact of the dead zone extent on the population of accreting transition disks. We discuss this point in \autoref{sec_Discussion_DeadZone}, and compare our approach with recent dead zone models.\par
Other mechanism angular momentum transport mechanisms, such as magnetic winds \citep[][]{Blandford1982} could lead to different disk accretion profiles, and it might be worth considering their effect on the global gas surface density evolution for future studies \citep[][]{Suzuki2016}.
\subsection{Dust evolution} \label{sec_Model_Dust}
Since we want to include a radiative transfer calculation in this work, we also need to include a model for the dust dynamics, since these differ from the dynamics of the gas \citep{Whipple1972, Weidenschilling1977}, as revealed by observations of disks at different wavelengths \citep[e.g. TW Hya,][]{Andrews2016, vanBoekel2017, Huang2018}.\par 
While small particles are indeed well coupled to the gas motion, larger particles can decouple from it, depending on their size.
A useful quantity to characterize the level of coupling is the Stokes number, $\textrm{St} = t_\textrm{stop} \Omega_k$, where $t_\textrm{stop}$ is the time necessary for the dust grains to couple to the gas motion due to their mutual drag force, and $\Omega_k$ is the Keplerian angular velocity.\par
For spherical dust grains located at the disk midplane, we can write the Stokes number of a particle of radius $a$, and material density $\rho_s$ as:
\begin{equation} \label{eq_StokesMidplane}
    \textrm{St} = \frac{\pi}{2}\frac{a\, \rho_s}{\Sigma_g} \cdot
    \begin{cases}
				1 & \lambda_\textrm{mfp}/a \geq 4/9\\
                \frac{4}{9} \frac{a}{\lambda_\textrm{mfp}} & \lambda_\textrm{mfp}/a < 4/9,
	\end{cases} 
\end{equation}
where $\lambda_\textrm{mfp} = (n \sigma_{\textrm{H}_2})^{-1}$ is the mean free path, with $n$ the number density and $\sigma_{\textrm{H}_2} = \SI{2e-15}{cm^2}$ the molecular cross section.\par
Because most of the mass is concentrated in large grains, and these tend to settle towards the midplane, \autoref{eq_StokesMidplane} is a convenient expression to describe the dust aerodynamic behavior.\par 
Following \cite{Nakagawa1986}  and \cite{Takeuchi2002}, the dust radial velocity is given by:
\begin{equation} \label{eq_dust_radial_velocity}
    v_\textrm{d} = \frac{1}{1 + \mathrm{St}^2} v_\textrm{g} -  \frac{2 \mathrm{St}}{1 + \mathrm{St}^2} \eta v_k,
\end{equation}
with $\eta = -\, (1/2)\,  (h_g / r)^2\, \partiallogdiff{P}$, $v_k$ the Keplerian orbital velocity, $P = \rho_\textrm{g,0} c_s^2 /\gamma$ the isothermal pressure, $h_g = c_s/\Omega_k$ the gas scale height and $\rho_\textrm{g,0}$ the gas volume density at the midplane.\par
It can also be useful to think of the Stokes number as a \dbquote{dynamical grain size}. From \autoref{eq_dust_radial_velocity} we see that small grains (with $\textrm{St} \ll 1$) move with the viscous velocity of the gas, large boulders ($\textrm{St} \gg 1$) do not move in the radial direction, and mid-sized pebbles ($\textrm{St} \approx 1$) drift towards the pressure maximum at $v_\textrm{d} \approx -\eta v_k$.\par
In addition to advection, dust particles also diffuse according to the concentration gradient, with a diffusivity of $D_\textrm{d} = \nu/(1 + \textrm{St}^2)$ \citep{Youdin2007}. The dust evolution is then described by the following advection-diffusion equation \citep{Birnstiel2010}:
\begin{equation} \label{eq_dust_advection}
    \partialdiff{t} \left(r \, \Sigma_{\textrm{d}}\right) + \partialdiff{r} (r \, \Sigma_{\textrm{d}} \, v_{\textrm{d}}) - \partialdiff{r} \left(r D_\textrm{d} \Sigma_\textrm{g} \partialdiff{r}\left(\frac{\Sigma_\textrm{d}}{\Sigma_\textrm{g}}\right)\right) = - \dot{\Sigma}_\textrm{w,d},
\end{equation}
where $\Sigma_\textrm{d}$ is the dust surface density (of a particular dust species), and $\dot{\Sigma}_\textrm{w,d}$ is the dust loss rate due to entrainment with the photoevaporative wind \citep{Hutchison2016, Franz2020, Hutchison2021, Booth2021}, which we describe below.\par
In this work we ignore the effects of the dust back-reaction onto the gas dynamics, since these become negligible at low dust-to-gas ratios ($\epsilon \leq 0.01$) and over long $\sim$ Myr timescales \citep{Garate2020}.
\subsubsection{Dust settling and wind entrainment}
Dust models predict that as particles grow in size, they also tend to settle towards the midplane \citep{Dubrelle1995}, a behavior that is confirmed by observations \cite[see][ for a prominent example]{Villenave2019, Villenave2020}.\par
Assuming that the gas is in hydrostatic equilibrium with a characteristic scale height $h_\textrm{g}$, the dust scale height can be approximated by \citep{Youdin2007}:
\begin{equation}\label{eq_scaleheight_dust}
h_\textrm{d} = h_\textrm{g} \cdot \min\left(1, \sqrt{\frac{\alpha}{\min(\mathrm{St},1/2) (1+\mathrm{St}^2)  }}\right).
\end{equation}\par
Then, the vertical structure of the gas and dust can be modeled with a Gaussian distribution \citep{Fromang2009} as:
\begin{equation} \label{eq_vertical_density}
\rho_\textrm{g,d}(z) = \frac{\Sigma_\textrm{g,d}}{\sqrt{2\pi} h_\textrm{g,d}} \exp\left(-\frac{z^2}{2 h^2_\textrm{g,d}}\right).
\end{equation}\par
Since small particles are well coupled to the gas and we expect the photoevaporative wind to be launched from the disk surface, we can estimate the dust loss rate due to wind entrainment as:
\begin{equation} \label{eq_dust_entrainment}
   \dot{\Sigma}_\textrm{w,d} = \epsilon_{a_w, h_w}  \dot{\Sigma}_\textrm{w},
\end{equation}
where  $\epsilon_{a_w, h_w}$ is the dust-to-gas ratio of small entrained particles ($a \leq a_w = \SI{10}{\mu m}$) above several gas scale heights ($z \geq h_w = 3 h_\textrm{g}$). We pick these limits based on the model of \cite{Franz2020}, which indicate that only small particles that are several scale heights above the midplane can get entrained with the wind.
While this approximation is simplistic, we find it effective because the dependency of the mass loss rate with the $a_w$ and $h_w$ parameters is very weak (G\'arate et al., in prep.).
\subsubsection{Dust growth} 
The last ingredients of our dust evolution model are growth and fragmentation which are necessary if we intend to model the simultaneous evolution of gas and dust \citep[e.g.,][]{Brauer2008, Birnstiel2010, Drazkowska2019}.\par
In this work, we consider that the dust phase consists of a distribution of particle species with different sizes, with their dynamics determined by their Stokes number (\autoref{eq_StokesMidplane}), that can also evolve through sticking and fragmentation, as described in \cite{Birnstiel2010}.\par
Without entering into the details of the coagulation model, we want to remark that the growth of a particle distribution is locally limited either by drift, when the drift timescale exceeds the growth timescale, or by fragmentation, when the collision velocity between particles exceeds the material fragmentation velocity $v_\textrm{frag}$ \citep{Ormel2007, Brauer2008, Birnstiel2009}.\par
Then, the maximum Stokes number that a particle can reach in each case is:
\begin{equation} \label{eq_drift_limit}
    \mathrm{St}_{\textrm{drift}} = \left|\frac{\textrm{dln}\, P}{\textrm{dln}\, r }\right|^{-1} \frac{v_k^2}{c_s^2} \epsilon,
\end{equation}
\begin{equation} \label{eq_fragmentation_limit}
    \mathrm{St}_{\textrm{frag}} = \frac{1}{3}\frac{v_\textrm{frag}^2}{\alpha c_s^2},
\end{equation}
with the maximum grain size given by $\mathrm{St}_\textrm{max} = \textrm{min}(\mathrm{St}_{\textrm{frag}}, \mathrm{St}_{\textrm{drift}})$ \citep[][]{Birnstiel2010, Birnstiel2012}.\par
Because the collision velocity between particles depends on the local turbulence parameter \citep{Ormel2007}, with:
\begin{equation} \label{eq_turbulent_speed}
    \Delta v_\textrm{turb} \approx \sqrt{\frac{3 \alpha}{\textrm{St} + \textrm{St}^{-1}}} c_s,
\end{equation}
we can expect particles to grow into larger sizes in regions with low $\alpha$ (see \autoref{eq_fragmentation_limit}), such as the dead zone \citep{Birnstiel2012_b, Pinilla2016, Ueda2019}.
\section{Simulation setup} \label{sec_Setup}
We perform our numerical simulations using the code \texttt{DustPy}
\footnote{A previous version of \texttt{DustPy} was used for this paper. The current version is available at \href{https://github.com/stammler/DustPy}{github.com/stammler/DustPy}.} (Stammler and Birnstiel, in prep.), 
that can simulate the advection of gas and dust, along with the growth and fragmentation of multiple particle species, following the model of \cite{Birnstiel2010}. The code also allows to load custom modules, which is how we implemented the photoevaporation and dead zone profiles described in \autoref{sec_Model}.\par
This study is separated into \dbquote{gas only} simulations, that are fast to run and ideal for population synthesis studies (which are the main focus of this work) and a single \dbquote{gas and dust} simulation, in which we solve the evolution of both components simultaneously to perform a radiative transfer calculation of a specific case of study, in order to compare with multi-wavelength observations.\par 
In this section, we describe the parameters used to set up the protoplanetary disk, the numerical grid and the parameter space explored.
\subsection{Disk setup} \label{sec_SetupDisk}
For our simulations we use a solar mass star, and  a circumstellar disk with an initial mass of $M_\textrm{disk} = 0.1 M_\odot$.
The initial gas surface density is defined with the \cite{Lynden-Bell1974} self-similar solution:
\begin{equation} \label{eq_LBPprofile}
    \Sigma_\textrm{g}(r) = \Sigma_0 \left(\frac{r}{r_c}\right)^{-1} \exp(-r/r_c).
\end{equation}
This initial condition is defined by the characteristic radius $r_c$ at which the exponential drop begins, and a normalization factor $\Sigma_0$, defined such that $\int 2\pi r \Sigma_\textrm{g} \,dr = M_\textrm{disk}$.\par
For the gas temperature we assume that the disk is heated passively by the central star, and use the following profile:
\begin{equation}
    T(r) = 150 \left(\frac{r}{\SI{1}{AU}} \right)^{-1/2} \SI{}{K}.
\end{equation}\par
For all our simulations, the radial grid extends from $\SI{1}{AU}$ to $\SI{300}{AU}$, with $n_r = 250$ logarithmically spaced grid cells.\par

Additionally, in \autoref{sec_Appendix_GasDust} we test the effect of the initial condition on our results, specifically, we present what would happen if the disk inner regions were in a quasi-steady state from the beginning of the simulation (i.e. a radially constant accretion profile, that decreases over time).
\subsection{Population synthesis} \label{sec_SetupPopulation}
\begin{table}
 \caption{Parameter space for dead zone models.}
 \label{Table_DZParam}
 \centering
  \begin{tabular}{ c  c }
    \hline \hline
    \noalign{\smallskip}
    Variable & Value  \\
    \hline
    \noalign{\smallskip}
    $\alpha_\textrm{a}$ &  $10^{-3}$ \\
    $\alpha_\textrm{dz}$ & [1, 3, 5]  $\times 10^{-4}$ \\
    $r_\textrm{dz}$ [AU] & [5, 10, 20]   \\
    \hline
  \end{tabular}
\end{table}
In this work we construct 10 different disk populations, consisting of one \dbquote{Control} population without a dead zone (i.e., $\alpha(r) = \alpha_\textrm{a}$), and 9 different \dbquote{Dead Zone} populations, with varying radial extents $r_\textrm{dz}$ and turbulence values $\alpha_\textrm{dz}$.
In \autoref{Table_DZParam} we list the parameter space explored for $r_\textrm{dz}$ and $\alpha_\textrm{dz}$. The active turbulence value of $\alpha_\textrm{a} = \SI{e-3}{}$ is kept constant for all the simulations.\par
To construct the disk populations, we use a similar approach to the previous studies from \cite{Owen2011, Ercolano2018} and \cite{Picogna2019}.
Each population consists of 1000 simulations, where the X-ray luminosity is sampled from the Taurus luminosity distribution, with values between $L_x = \SI{e28}{erg\, s^{-1}}$ and $ L_x =\SI{e31}{erg\, s^{-1}}$ \citep{Preibisch2005, Gudel2007}, and the disk characteristic size is sampled from a uniform distribution with values between $r_c = \SI{20}{AU}$ and $r_c = \SI{100}{AU}$.
To ensure that the populations are comparable between each other, we use the same sample of $L_x$ and $r_c$ across the different populations.\par
We track the evolution of each simulation for $\SI{10}{Myr}$, or until the photoevaporation carves a gap that extends beyond $\SI{100}{AU}$, saving a snapshot every $\SI{0.1}{Myr}$
\footnote{The simulation data from the population synthesis and plotting routines are available in Zenodo: \href{https://doi.org/10.5281/zenodo.4761432}{doi.org/10.5281/zenodo.4761432}}.\par 
In particular we want to obtain the probability distribution of the gas accretion rate $\dot{M}_\textrm{g}$ (measured at \SI{1}{AU}) and the outer edge of the gap opened by photoevaporation including the presence of a dead zone, for each population model (see \autoref{sec_Results_Gas}), and compare them to the observed distribution of accreting transition disks.
\subsection{Model with dust evolution}\label{sec_SetupDust}
We perform an additional simulation including dust evolution (as described in \autoref{sec_Model_Dust}), and using the following parameters: $L_x = \SI{e31}{erg\, s^{-1}}$, $r_c = \SI{60}{AU}$, $\alpha_\textrm{dz} = \SI{e-4}{}$, and $r_\textrm{dz} = \SI{5}{AU}$. The remaining parameters are the same as in \autoref{sec_SetupDisk}.\par
For the size distribution we consider a logarithmic grid of $n_m = 141$ particle species going from $\SI{5e-5}{cm}$ to $\SI{2.5e2}{cm}$. The initial dust size distribution follows the \cite{Mathis1977} power-law, with a maximum size of $a_0 = \SI{1}{\mu m}$, and we assume an initial dust-to-gas ratio of $\epsilon_0 = 0.01$.\par
For the dust properties we assume that the fragmentation velocity is $v_\textrm{frag} = \SI{10}{m\, s^{-1}}$ \citep{Wada2011, Gundlach2011, Gundlach2015}, that the grains have a material density of $\rho_s = \SI{1.6}{g\, cm^{-3}}$ and that these are compact, with the grain mass given by $m = 4\pi \rho_s a^3/3$.\par
The dust then evolves according to the growth and advection model described in \autoref{sec_Model_Dust}. Finally, we can use the dust size distribution to obtain the disk's SED and its brightness in both the millimeter continuum ($\SI{1.3}{mm}$) and scattered light ($\SI{1.25}{\mu m}$), using the radiative transfer code \texttt{RADMC-3D}\footnote{\href{https://www.ita.uni-heidelberg.de/~dullemond/software/radmc-3d/}{www.ita.uni-heidelberg.de/~dullemond/software/radmc-3d/}} \citep{RADMC3D2012}.\par
We restricted our analysis to a single \dbquote{dust and gas} simulation for two reasons:
First, because these simulations are computationally expensive to run.
The second reason is that we are particularly interested in the scenario where a large gap opens on short timescales ($\approx \SI{1}{Myr}$), while keeping the presence of a compact inner dust disk, as this has been seen in observations of transition disks at different wavelengths \citep[e.g.][]{Benisty2010, Olofsson2013, Perraut2019_Gravity, Pinilla2019, Bouarour2020_Gravity, Francis2020, Rosotti2020}. 
In order to ensure the fast gap opening, we use a high value for the X-ray luminosity ($L_x = \SI{e31}{erg\, s^{-1}}$) in this simulation.\par
We defer an in-depth study of the dust evolution to a follow up work, and limit ourselves to comment on the expected effect of the existing parameters based on the available results.
%
\section{Results - Gas evolution} \label{sec_Results_Gas}
In this section we present the results of the population synthesis study with gas only simulations, showing that disks with dead zones are more likely to continue accreting during photoevaporative dispersal, than those without.\par
As a simplification, we will refer to a disk as \dbquote{accreting} if the accretion rate at \SI{1}{AU}  is $\dot{M}_\textrm{g} \geq \SI{e-11}{M_\odot yr^{-1}}$, and \dbquote{non-accreting} otherwise. We pick this limit in order to be consistent with previous works \citep[e.g.][]{Owen2011}, and observational detection limits \citep{Alcala2017}.
\subsection{Proof of concept}
\begin{figure}
\centering
\includegraphics[width=90mm]{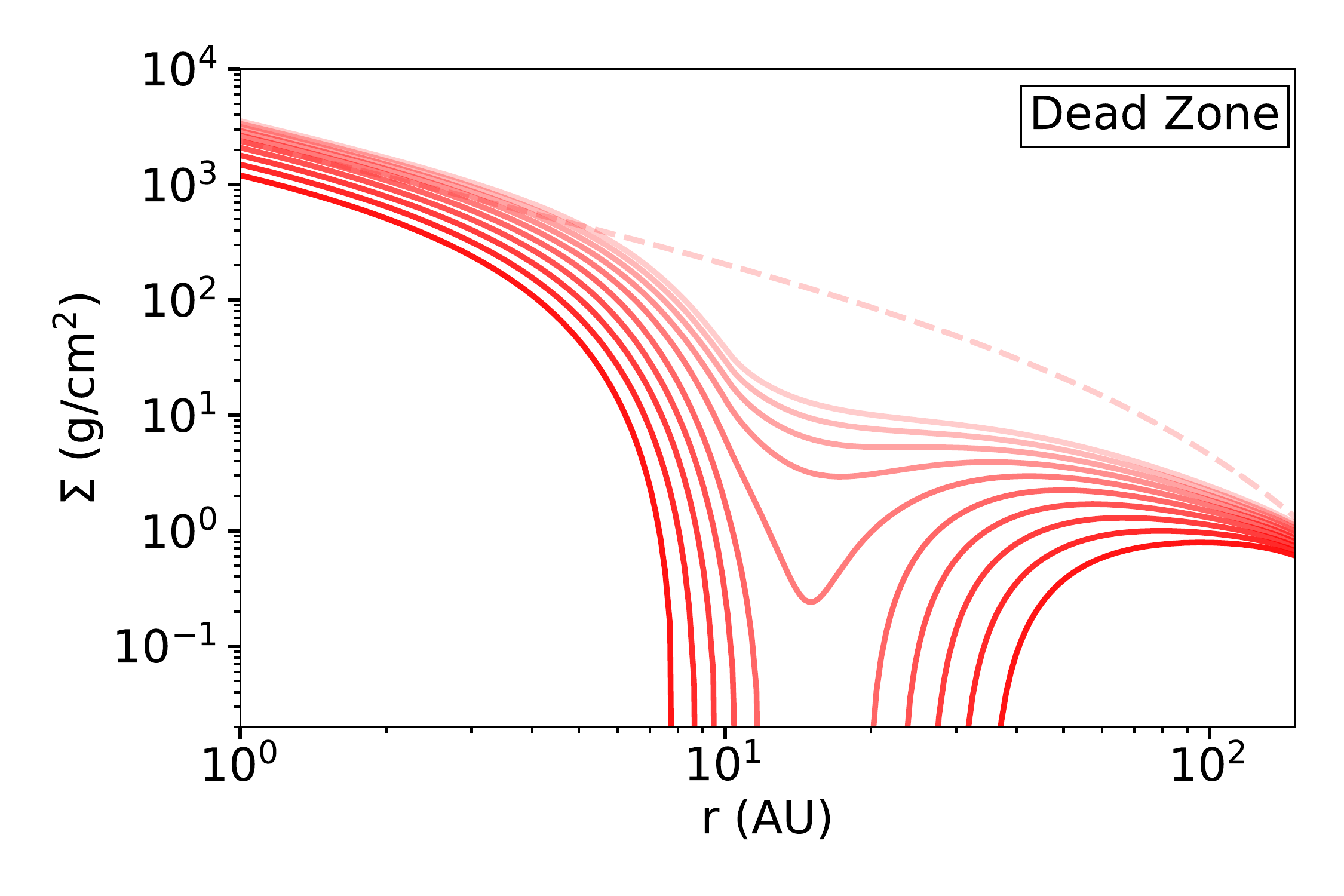}
\includegraphics[width=90mm]{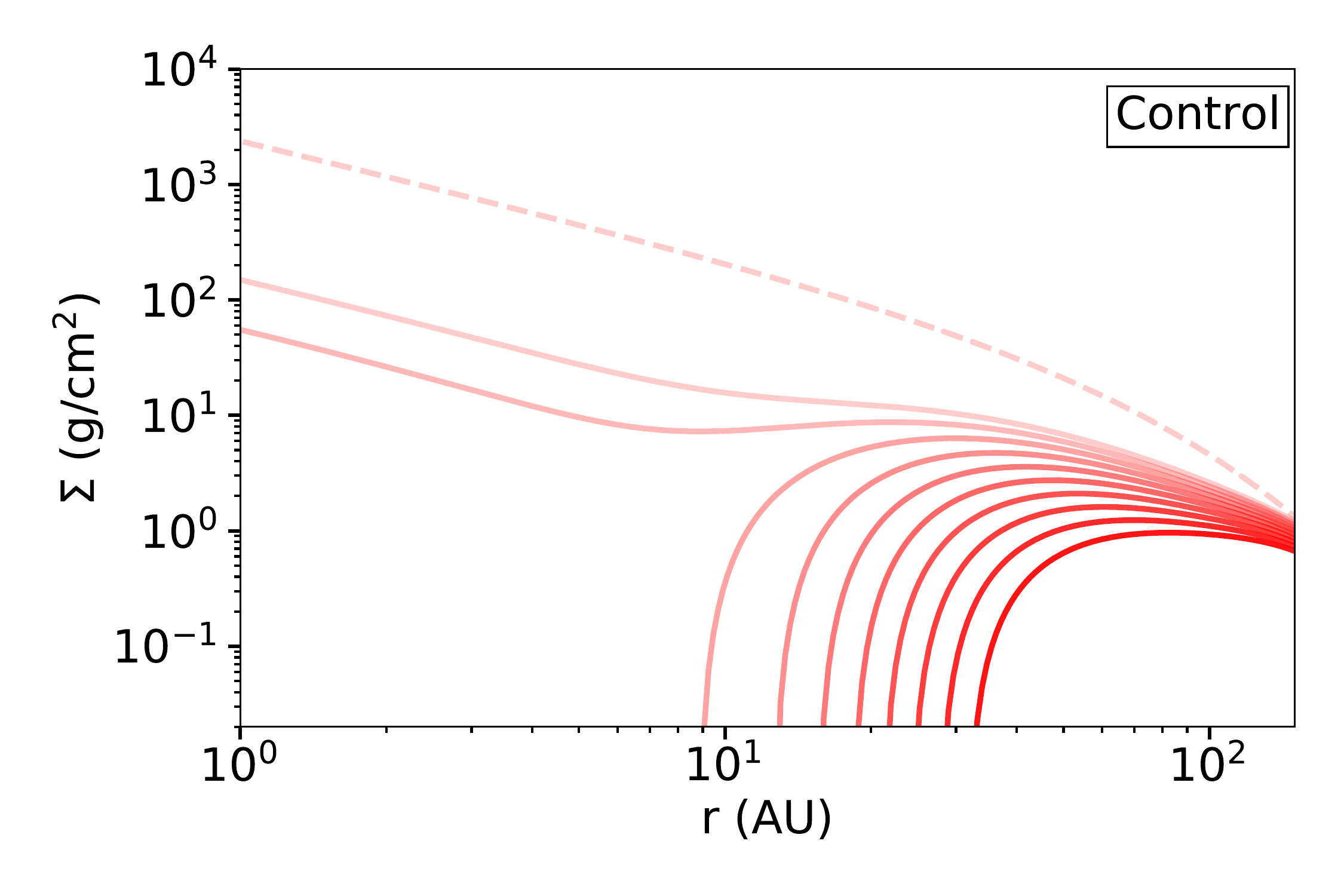}
 \caption{
 Gas surface density evolution of two simulations with (top) and without (bottom) a dead zone, undergoing photoevaporative dispersal (with $L_x = \SI{e30}{erg\, s^{-1}}$). The dashed line shows the initial gas surface density. The solid lines show the gas evolution every \SI{0.2}{Myr}, starting at \SI{2.5}{Myr}, where the opacity increases with time. The dead zone parameters for the top panel are $\alpha_\textrm{dz} = \SI{e-4}{}$ and $r_\textrm{dz} = \SI{10}{AU}$.
}
 \label{Fig_SurfaceDensityEvolution}
\end{figure}
\begin{figure}
\centering
\includegraphics[width=90mm]{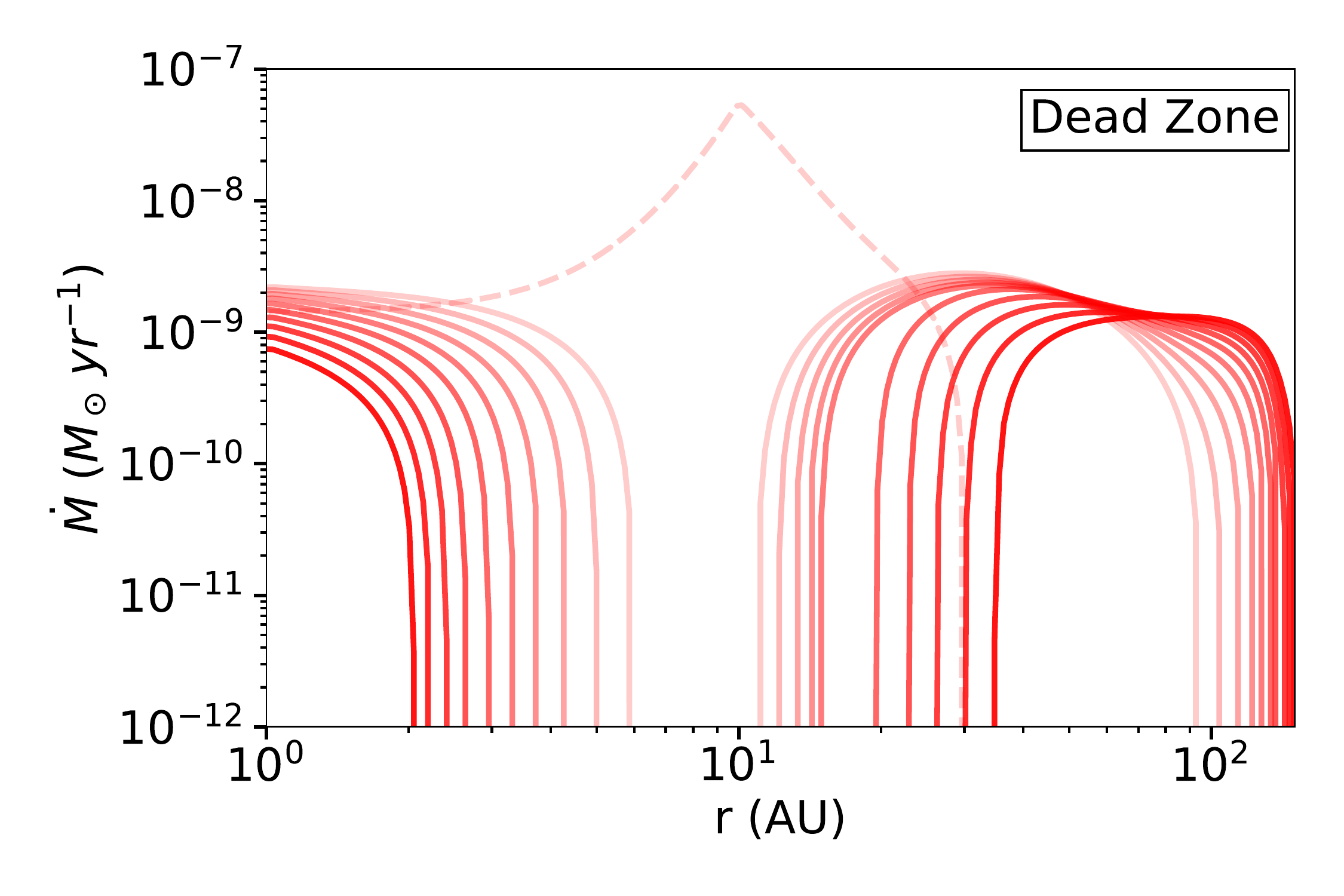}
\includegraphics[width=90mm]{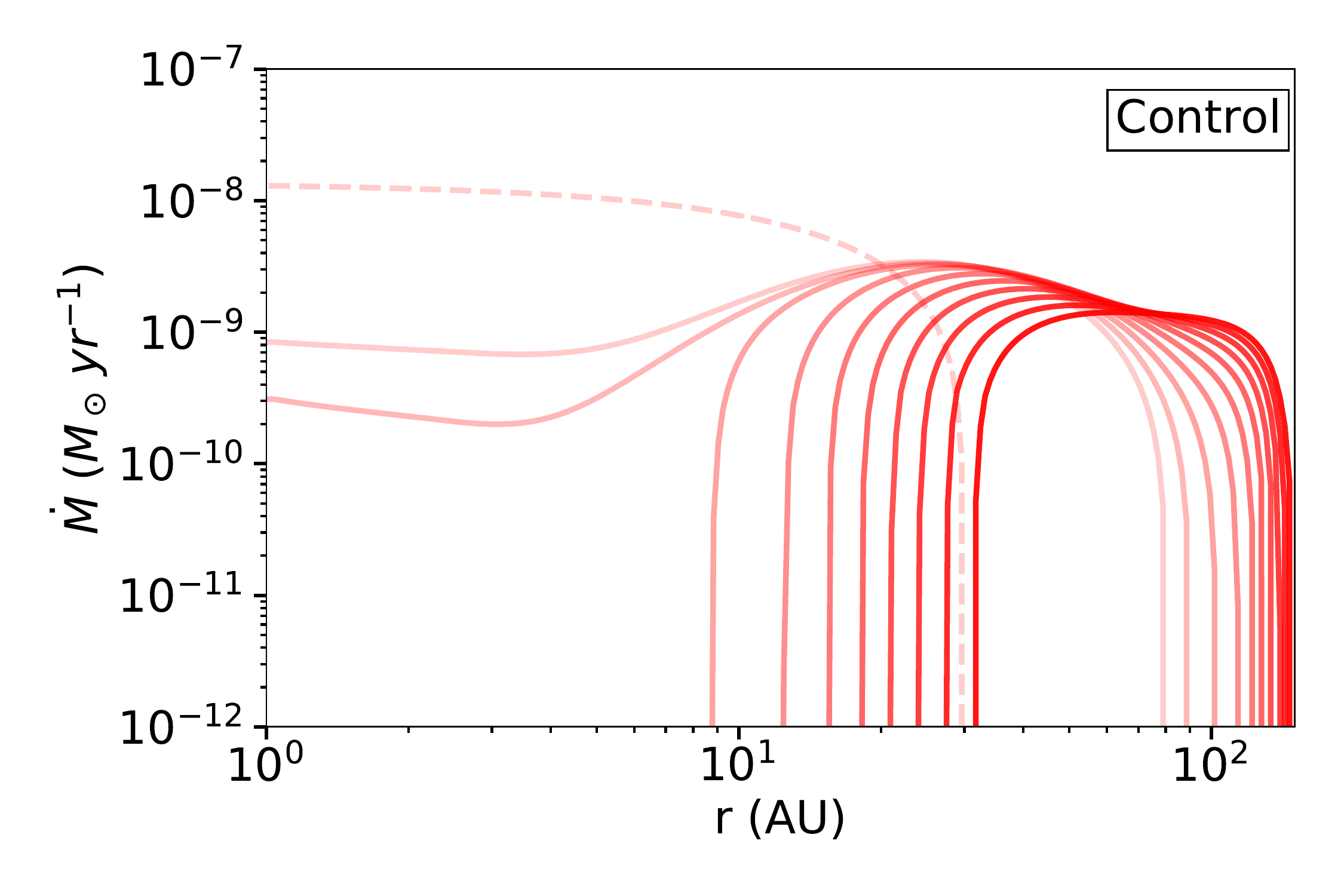}
 \caption{
 Accretion rate profiles for two simulations with (top) and without (bottom) a dead zone. The dashed line shows the initial accretion profile. The solid lines show the gas evolution every \SI{0.2}{Myr}, starting at \SI{2.5}{Myr}, where the opacity increases with time. 
 }
 \label{Fig_AccretionEvolution}
\end{figure}
To first illustrate how dead zones allow for a dispersing disk to sustain a high accretion rate we present the gas surface density evolution of two disks (with and without a dead zone) in \autoref{Fig_SurfaceDensityEvolution}.
Both disks were constructed following \autoref{sec_SetupDisk}, with $r_c = \SI{60}{AU}$ and $L_x = \SI{e30}{erg\, s^{-1}}$. For the disk with a dead zone, we used an extent of $r_\textrm{dz} = \SI{10}{AU}$ and turbulence of $\alpha_\textrm{dz} = \SI{e-4}{}$ (see \autoref{Fig_AlphaModel}).\par
\autoref{Fig_SurfaceDensityEvolution} shows how the inner and outer parts of the disk with a dead zone evolves on different timescales. The outer disk is dispersed once the local accretion rate drops below the photoevaporation mass loss rate \citep{Clarke2001}, while the inner disk evolves on a longer viscous timescale \citep{Morishima2012}.\par
In contrast, the inner region of the disk without a dead zone gets accreted onto the star immediately after the gap opens.\par
\autoref{Fig_AccretionEvolution} shows how the disk with a dead zone maintains an accretion rate of $\dot{M}_\textrm{g} \approx \SI{e-9}{M_\odot\, yr^{-1}}$ during its entire evolution, even after the inner disk is disconnected from the outer region. On the other hand, for the model without a dead zone the accretion rate drops quickly after the gap opens, turning it into a non-accreting disk.
\subsection{Population synthesis models}
\begin{figure}
\centering
\includegraphics[width=95mm]{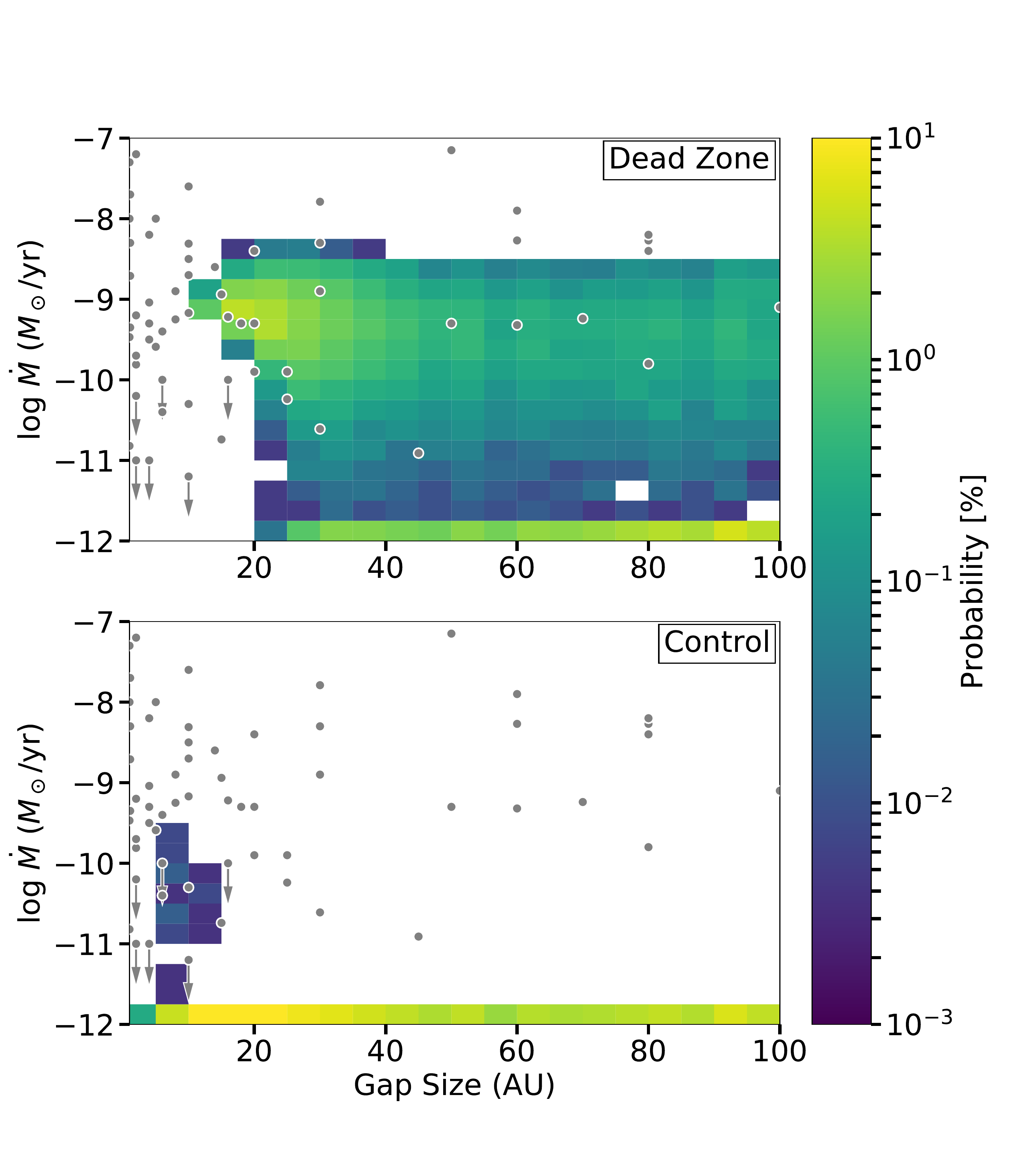}
 \caption{Probability distribution of the gap size (measured with its outer edge) and accretion rate (measured at 1.5 AU) for a population synthesis model with (top) and without (bottom) a dead zone. The dead zone model uses $r_\textrm{dz}= \SI{10}{AU}$ and $\alpha_\textrm{dz} = \SI{e-4}{}$.
 Disks with accretion rates lower than $\SI{e-12}{M_\odot\, yr^{-1}}$ are counted in the lowest accretion rate bin.
 The gray dots mark the observed population of transition disks compiled in the \cite{Ercolano2017_Review} review. The dots with arrows mark upper limits in the accretion rate.
 }
 \label{Fig_ProbabilityDist}
\end{figure}

\begin{figure}
\centering
\includegraphics[width=90mm]{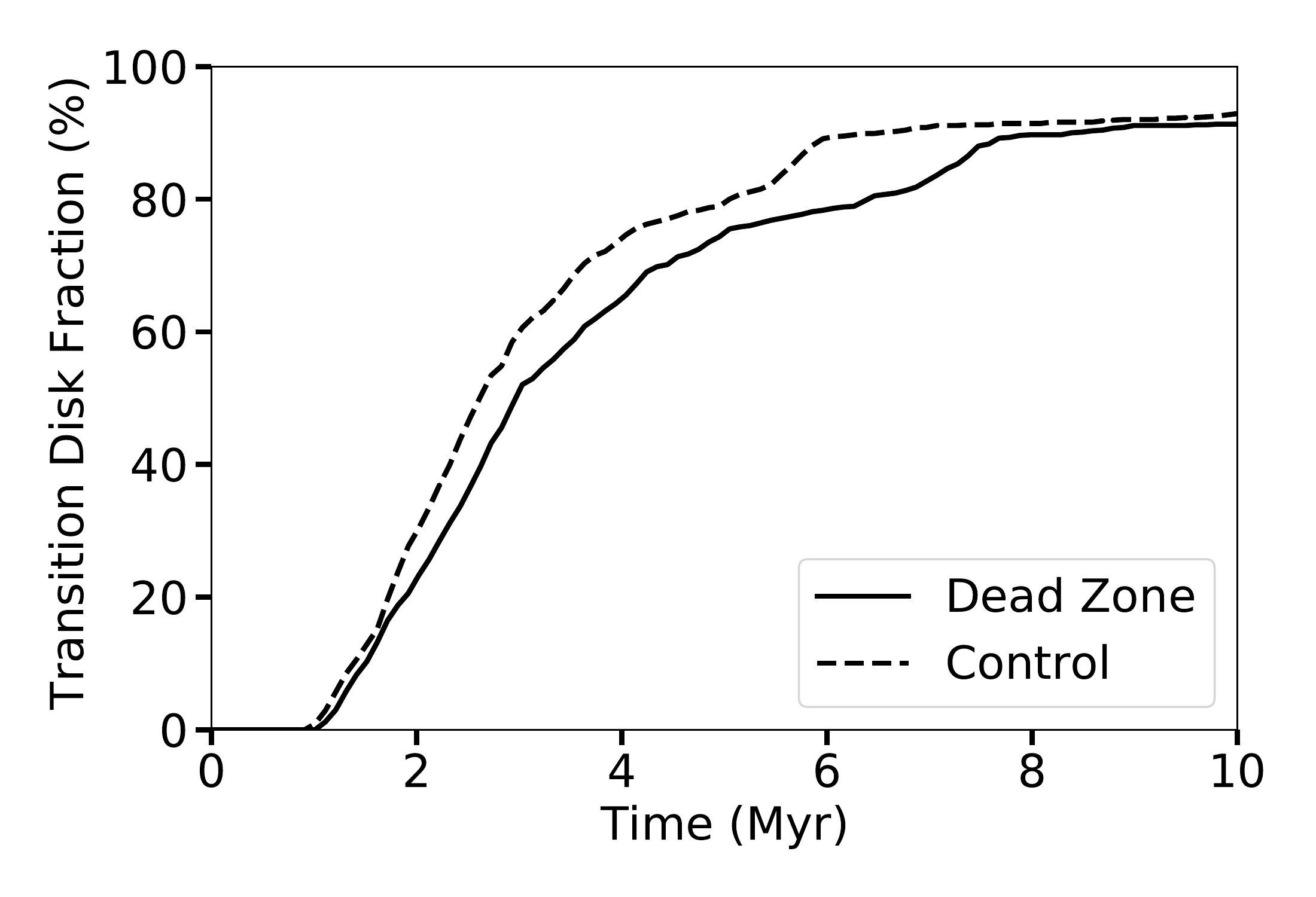}
 \caption{
 Evolution of the fraction of transition disks, relative to the whole disk population, over time for both the \dbquote{Control} and \dbquote{Dead Zone} populations shown in \autoref{Fig_ProbabilityDist}. For simplicity, we identify a \dbquote{transition disk} as a disk with a gap opened by photoevaporation.
 }
 \label{Fig_TransitionFraction}
\end{figure}


The 2D histograms in \autoref{Fig_ProbabilityDist} show the probability distribution of the accretion rate and gap size for two different population models of transition disks (each generated from 1000 simulations and sampling every $\SI{0.1}{Myr}$ from the moment a gap opens, see \autoref{sec_SetupPopulation}).\par
The \dbquote{Control} population, which consists of disks without a dead zone, shows that only a small fraction ($<1\%$) of the recorded transition disks are still accreting once the gap opens (with $\dot{M}_\textrm{g} > \SI{e-11}{M_\odot\, yr^{-1}}$). 
Also, once the gap exceeds \SI{20}{AU} in size, all transition disks have stopped accreting. The previous population studies of \cite{Owen2011} and \cite{Picogna2019} also show similar results.\par
The \dbquote{Dead Zone} population, on the other hand, shows that a high fraction of transition disks are accreting, with $31\%$ of them accreting at rates of $\dot{M}_\textrm{g} > \SI{5.e-10}{M_\odot\, yr^{-1}}$ and $63\%$ at rates of $\dot{M}_\textrm{g} > \SI{e-11}{M_\odot\, yr^{-1}}$. 
We also find that the fraction of accreting transition disks is higher in disks with smaller gaps. In particular, if we only look at transition disks with gaps sizes smaller than $\SI{50}{AU}$, we find that $85\%$ of them are accreting.\par
In \autoref{Fig_ProbabilityDist} we also over-plot the measured gap sizes and accretion rates of observed transition disk population \citep{vanderMarel2016, Keane2014, Manara2014, Manara2016_b, Manara2016_a, Manara2017, GarciaLopez2006, Pascucci2007, Najita2007, Spezzi2008, Merin2010, Donehew2011, Curran2011, Alcala2014, Cieza2010, Cieza2012, Romero2012, Follette2015}, using the objects listed in \citet[][Table A.1]{Ercolano2017_Review}.\par 
Comparing our results with the observed transition disks, we find that the model including a dead zone can reproduce the measured accretion rates and gap sizes larger than \SI{10}{AU}. However, we also notice that this model does not predict disks with gaps smaller than the dead zone radial extent $r_\textrm{dz}$, since the inner regions evolve too slowly for photoevaporation to carve a gap during the early stages of disk dispersal. 
In \autoref{sec_Discussion_RelicDisks} we discuss how disks with narrow gaps could still be predicted within our framework.\par
In \autoref{Fig_TransitionFraction} we show the fraction of transition disks relative to the whole disk population (i.e. transition and primordial disks) at each time. In both the \dbquote{Control}  and \dbquote{Dead Zone} population, we find that photoevaporation opens a gap in approximately $20\%$ of the disks by \SI{2}{Myrs}, and in $90\%$ of the disks at some point of their lifetime (i.e. within \SI{10}{Myrs}), indicating that while the dead zone helps to sustain the inner disk for longer timescales, it does not affect whether or not a gap opens, and this would be solely determined by the star's X-ray luminosity and the disk global properties (e.g. mass, size, and turbulence).
\subsection{Effect of the dead zone properties.}
\begin{figure*}
\centering
\includegraphics[trim={160px 75px 270px 135px},clip,width=0.99\textwidth]{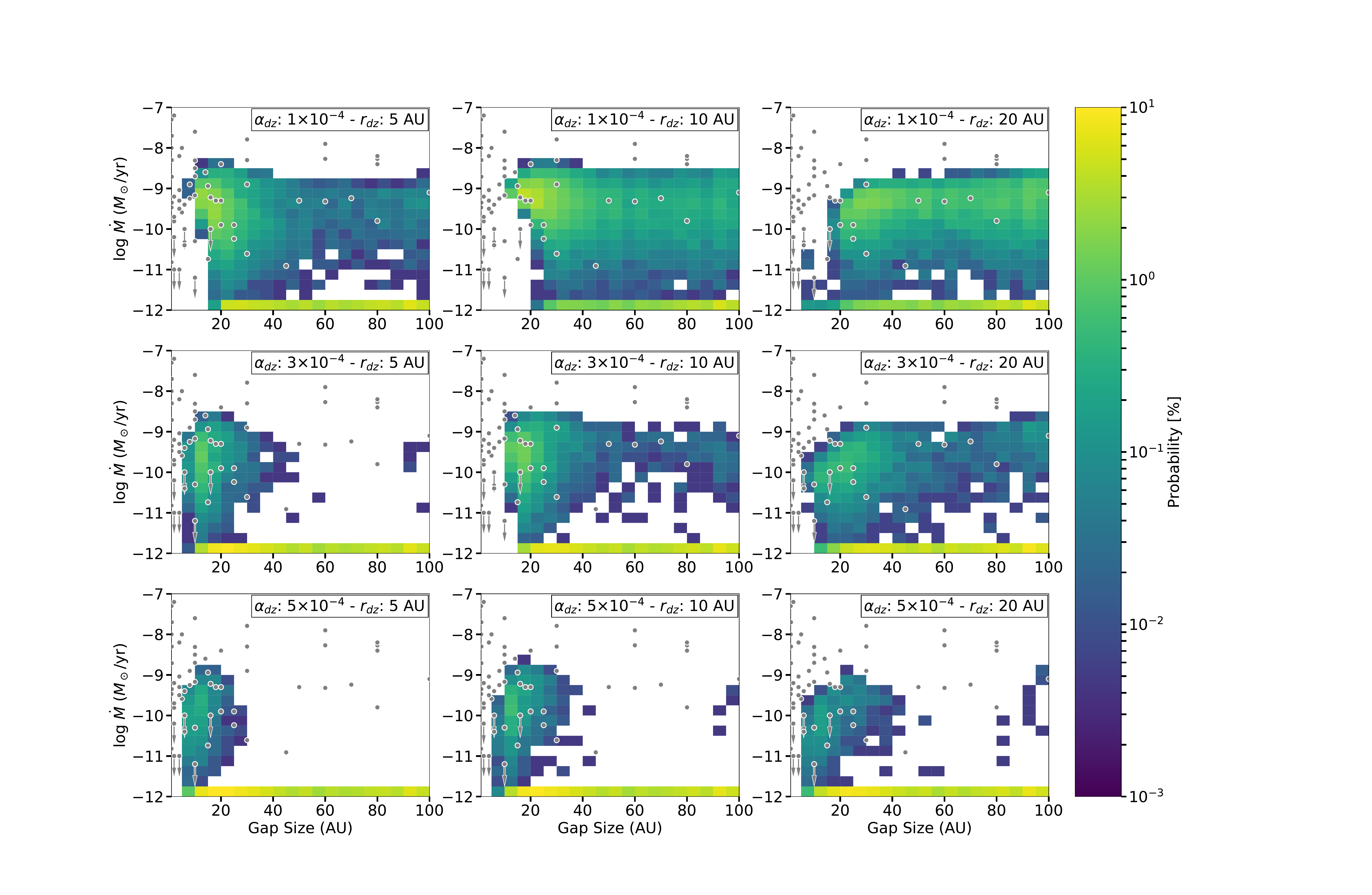}
 \caption{
 Same as \autoref{Fig_ProbabilityDist} for different dead zone models. The labels indicate the $\alpha_\textrm{dz}$ and $r_\textrm{dz}$ parameters used in each model.
 }
 \label{Fig_ProbabilityDist_Param}
\end{figure*}
\begin{table}
 \caption{Percentage of transition disks with $\dot{M}_g > \SI{e-11}{M_\odot\, yr^{-1}}$ for different dead zone models.}
 \label{Table_DiskFraction}
 \centering
  \begin{tabular}{ c | c | c}
    \hline \hline
    \noalign{\smallskip}
    $\alpha_\textrm{dz}$ & $r_\textrm{dz}$ (AU) & Accreting Disk Fraction (\%)\\
    \hline
    \noalign{\smallskip}
     & \SI{20}{} & 55.2 \\
    \SI{e-4}{} & \SI{10}{} & 62.9 \\
     & \SI{5}{} & 34.1  \\
    \hline
    \noalign{\smallskip}
     & \SI{20}{} & 12.3 \\
    \SI{3e-4}{} & \SI{10}{} & 16.9 \\
     & \SI{5}{} & 8.9  \\
    \hline
    \noalign{\smallskip}
     & \SI{20}{} & 2.4 \\
    \SI{5e-4}{} & \SI{10}{} & 4.6 \\
     & \SI{5}{} & 3.0  \\
    \hline
    \noalign{\smallskip}
    \dbquote{Control} & -- & 0.1\\
    \hline
  \end{tabular}
\end{table}
To study how the properties of the dead zone affect the population of transition disks, we conduct a parameter space study for different dead zones sizes ($r_\textrm{dz}$) and turbulence parameters ($\alpha_\textrm{dz}$), shown in \autoref{Fig_ProbabilityDist_Param}. To complement the figure we include the fraction of transition disks that present accretion rates higher than $\SI{e-11}{M_\odot\, yr^{-1}}$ in \autoref{Table_DiskFraction}, for the different dead zone properties.
In all cases we see a higher fraction of accreting transition disks when a dead zone is considered. Even the compact dead zone population with high turbulence ($\alpha_\textrm{dz} = \SI{5e-4}{}$ and $r_\textrm{dz}=\SI{5}{AU}$) shows $\approx 20$ times more accreting disks than the \dbquote{Control} population without a dead zone.\par
From the distribution of accretion rates and gap sizes we learn that disks with low turbulence ($\alpha_\textrm{dz} = \SI{e-4}{}$) present the higher fraction of accreting transition disks, which is expected since the inner disk will evolve on longer viscous timescales.\par
The effect of the dead zone size is less straightforward, with medium sized dead zones ($r_\textrm{dz} = \SI{10}{AU}$)  producing the highest fraction of accreting transition disks for a fixed $\alpha_\textrm{dz}$ (see \autoref{Table_DiskFraction}).
From our intuition we would expect larger dead zones to have longer viscous evolution timescales, with $t_\nu \sim r_\textrm{dz}^2 / \nu\, (r_\textrm{dz})$, and therefore to survive for longer, which is indeed what happens when we increase the dead zone size from \SI{5}{AU} to \SI{10}{AU}. 
However, when further increasing the dead zone size from \SI{10}{AU} to \SI{20}{AU} we find that the fraction of accreting disks decreases from 63\% to 55\% (for $\alpha_\textrm{dz} = \SI{e-4}{}$).
We can explain this behavior if we consider that before photoevaporation opens a gap, dead zones evolve together with the whole disk, receiving material from the outer regions which is then redistributed across the inner regions within the local viscous timescale. 
In mid-size dead zones the material is redistributed more efficiently than in more extended ones, resulting in a higher and spatially uniform accretion rate across the inner disk, and a longer lifetime. 
In contrast, for more extended dead zones the incoming material from the outer disk stays concentrated close around the dead zone outer edge ($r_\textrm{dz}$) for timescales of several million years. The innermost region then evolves isolated from the outer disk, allowing photoevaporation to open a gap within the dead zone itself, and reducing the fraction of accreting transition disks.
\par
Additionally, we also find that the fraction of accreting transition disks can be higher, if the initial mass distribution follows a quasi-steady state solution (see \autoref{sec_Appendix_GasDust}).

\section{Results - Dust evolution} \label{sec_Results_Dust}
\begin{figure}
\centering
\includegraphics[width=90mm]{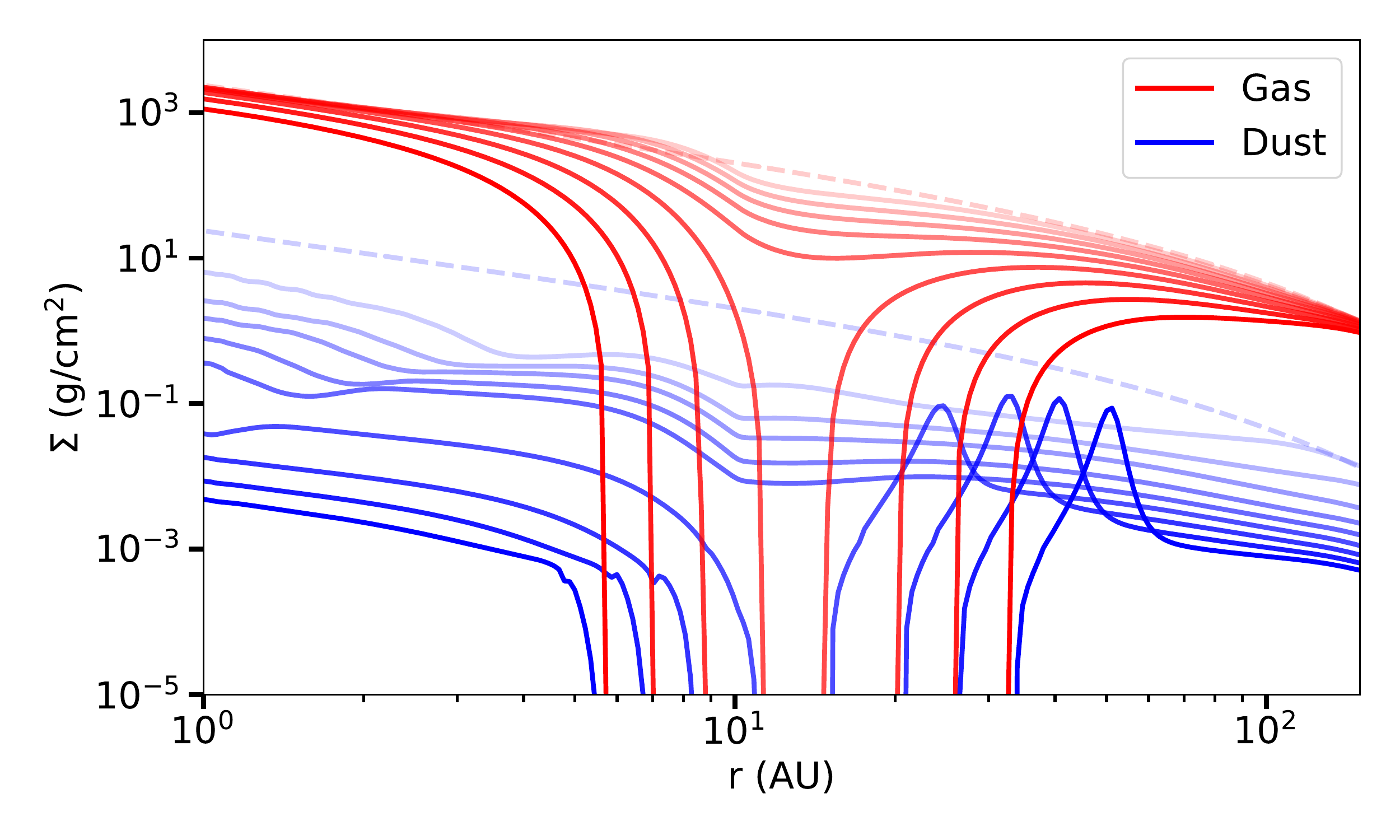}
 \caption{
 Surface density evolution for gas and dust (integrated for all particle sizes) in a disk with a dead zone (as described in \autoref{sec_SetupDust}). The dashed line shows the initial condition, the solid lines show the disk evolution every $\SI{0.2}{Myr}$. The star has an X-ray luminosity of $L_x = \SI{e31}{erg\, s^{-1}}$, which leads to a gap opening at $t \approx \SI{1}{Myr}$.
 }
 \label{Fig_SurfaceDensity_Dust}
\end{figure}

\begin{figure}
\centering
\includegraphics[width=95mm]{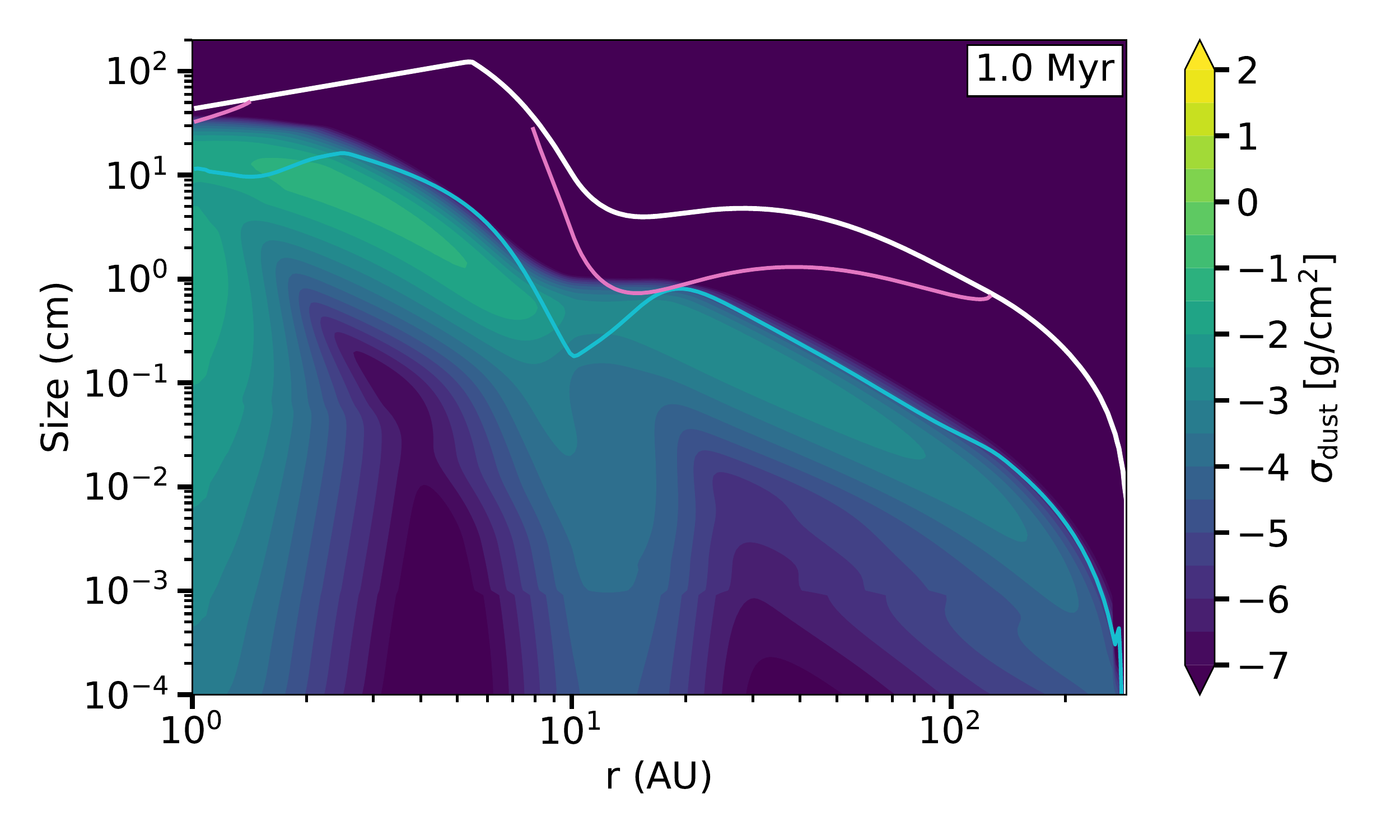}
\includegraphics[width=95mm]{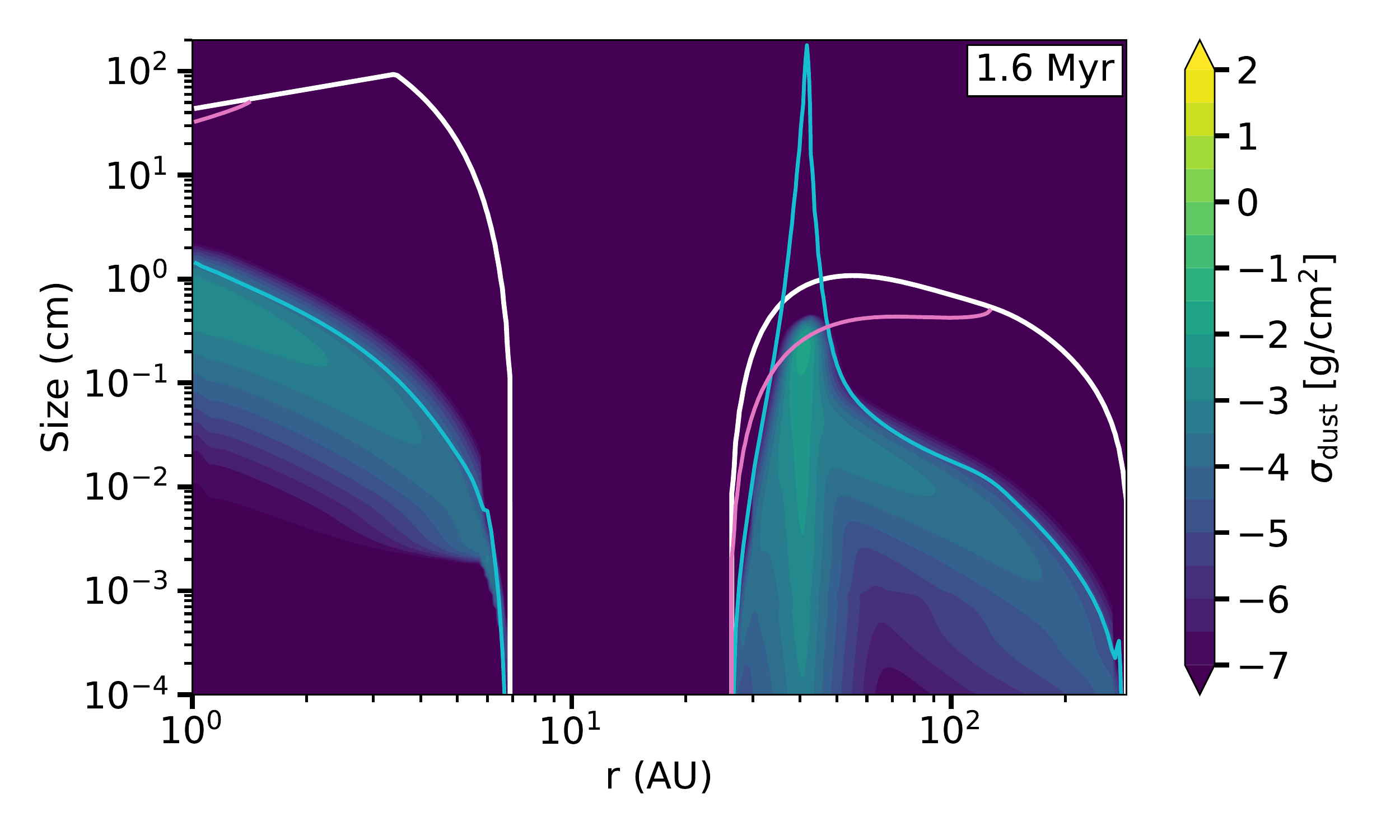}
\includegraphics[width=95mm]{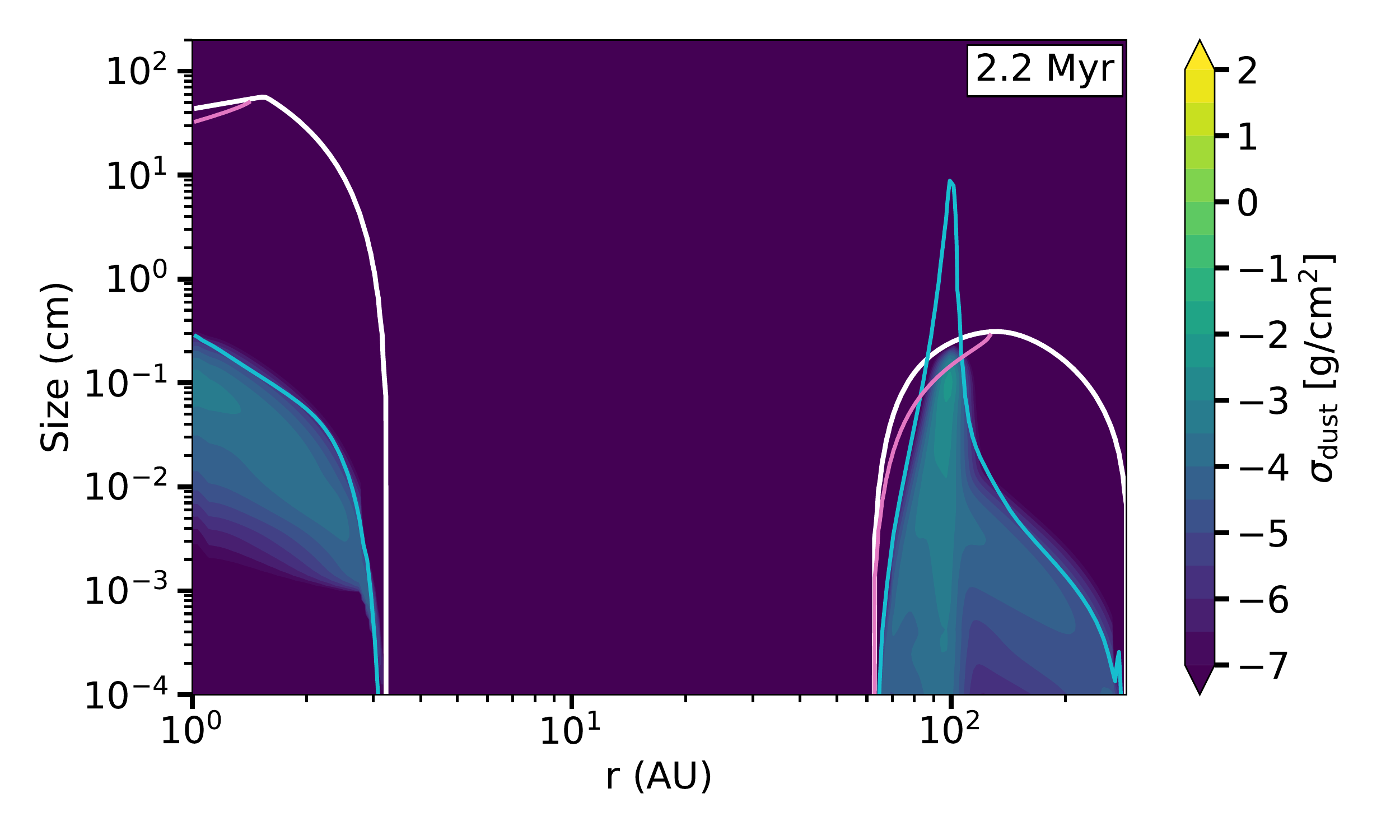}
 \caption{
 Dust size distribution for a disk with a dead zone (as described in \autoref{sec_SetupDust}) at \SI{1.0}{Myr}, \SI{1.6}{Myr}, and \SI{2.2}{Myr}. The solid lines show the drift growth limit (cyan), the fragmentation growth limit (pink), and the size corresponding to $\mathrm{St} = 1$ (white).
 }
 \label{Fig_DustSizeDist}
\end{figure}
In this section, we describe the evolution of the dust component for a disk with a dead zone (following the setup from \autoref{sec_SetupDust}), and show the expected emission using radiative transfer calculations.
\subsection{Dust distribution} \label{sec_Results_DustDistribution}
\autoref{Fig_SurfaceDensity_Dust} shows how the
dust surface density (integrated for all the grain sizes) evolves along with the gas during the disk dispersal in the case of a strong X-ray luminosity ($L_x = \SI{e31}{erg\, s^{-1}}$).\par
Before the gap opening, the dust drifts from the outer disk towards the inner dead zone regions, as it looses angular momentum due to the drag force from the gas \citep{Whipple1972, Weidenschilling1977}.\par
Because of the low turbulence in the dead zone region, dust particles can grow to larger sizes and therefore drift faster, depleting the inner disk of solids on shorter timescales (\autoref{eq_dust_radial_velocity} and \ref{eq_drift_limit}).\par
The dust size distribution (\autoref{Fig_DustSizeDist}, top panel) shows that in the dead zone the dust particles can grow up to sizes of $a \sim \SI{10}{cm}$, while in the outer disk we find mostly millimeter and sub-millimeter grains.\par
Additionally, we notice that our dead zone model results in a disk where particle growth is completely limited by drift, whereas disks without dead zones would be typically dominated by drift only in the outer regions ($r \gtrsim \SI{10}{AU}$ for $\alpha = \SI{e-3}{}$, after $\SI{1}{Myr}$), and by fragmentation in the inner ones \citep{Birnstiel2010, Birnstiel2012}.\par
Once the photoevaporation opens a gap (approximately at $\SI{1}{Myr}$), the inner and outer disk become disconnected in terms of mass transfer.\par
The inner disk continues to lose material as the dust drifts inward, though we also observe that the drift timescales increase over time. This happens because when the dust-to-gas ratio decreases, the maximum Stokes number decreases as well (\autoref{eq_drift_limit}), which leads to slower drift velocities (\autoref{eq_dust_radial_velocity}). 
This behavior is in line with the arguments described in \cite{Birnstiel2012, Powell2017}, in which the lifetime of disks with grain growth limited by drift is approximately equal to the growth and drift timescales. 
Since the growth timescale is approximately $t_\textrm{growth} = (\epsilon \Omega_k)^{-1}$, we find that the time elapsed after the gap opening (i.e. the lifetime of the inner disk as an isolated system) determines the dust-to-gas ratio present in the inner regions.\par
From the dust distribution of the inner region (\autoref{Fig_DustSizeDist}, middle and bottom panels), we observe that it is dominated mostly by large cm-sized grains, and that there is a lack of micron sized dust. This happens because the collision velocities in the dead zone are low (\autoref{eq_turbulent_speed}), and prevent the small particles to be replenished by fragmentation.\par
Meanwhile, the solids in the outer disk drift towards the edge of the photoevaporative gap, which is the new local pressure maximum.
The dust grains form a ring, in which growth is limited by fragmentation, reaching sizes of $a \approx \SI{1}{mm}$. As the gap expands over time, the dust ring moves along with it \citep{Alexander2007, Owen2019}.
\subsection{Dust emission} \label{sec_Results_DustEmission}
\begin{figure}
\centering
\includegraphics[width=95mm]{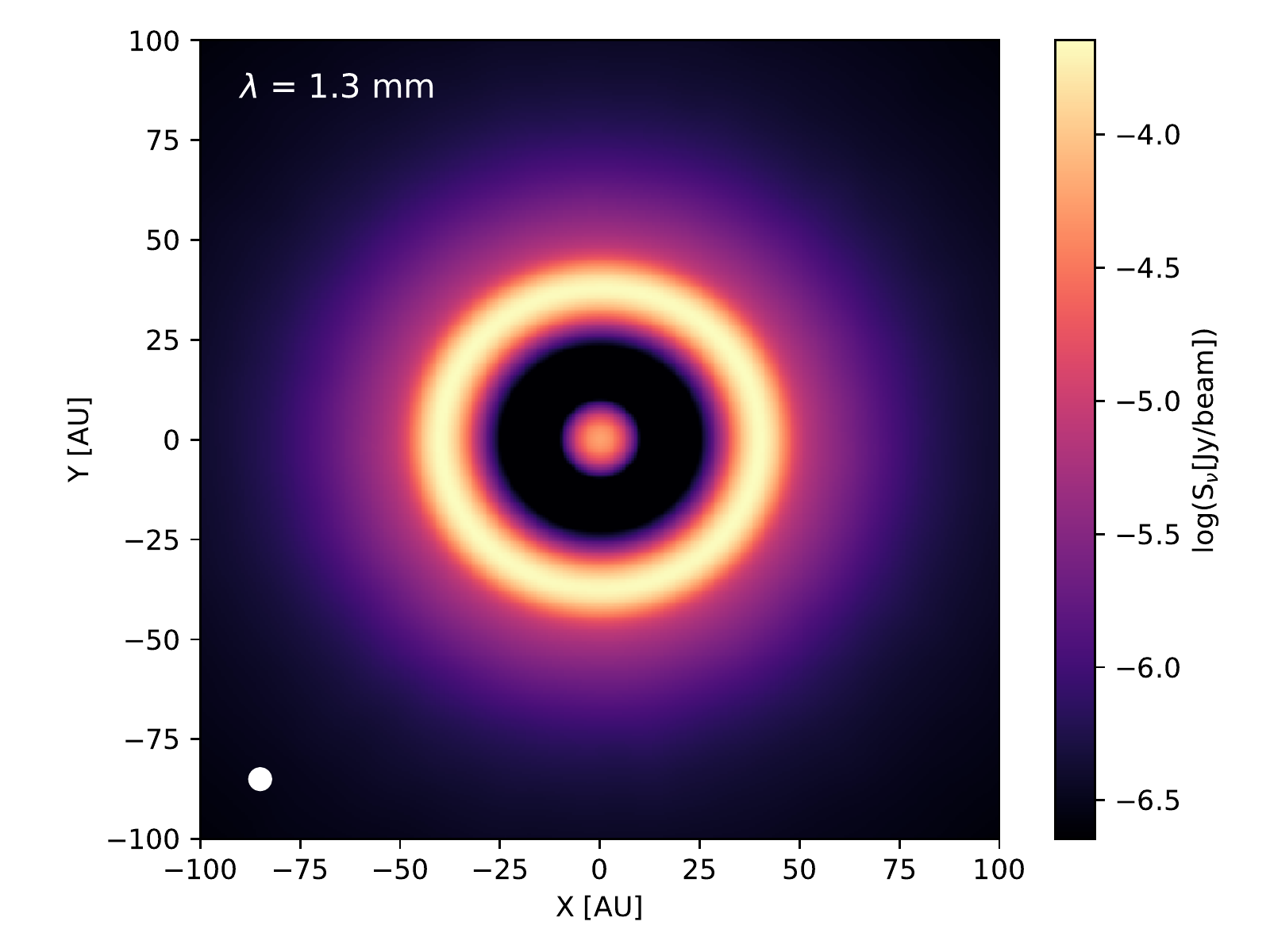}
\includegraphics[width=95mm]{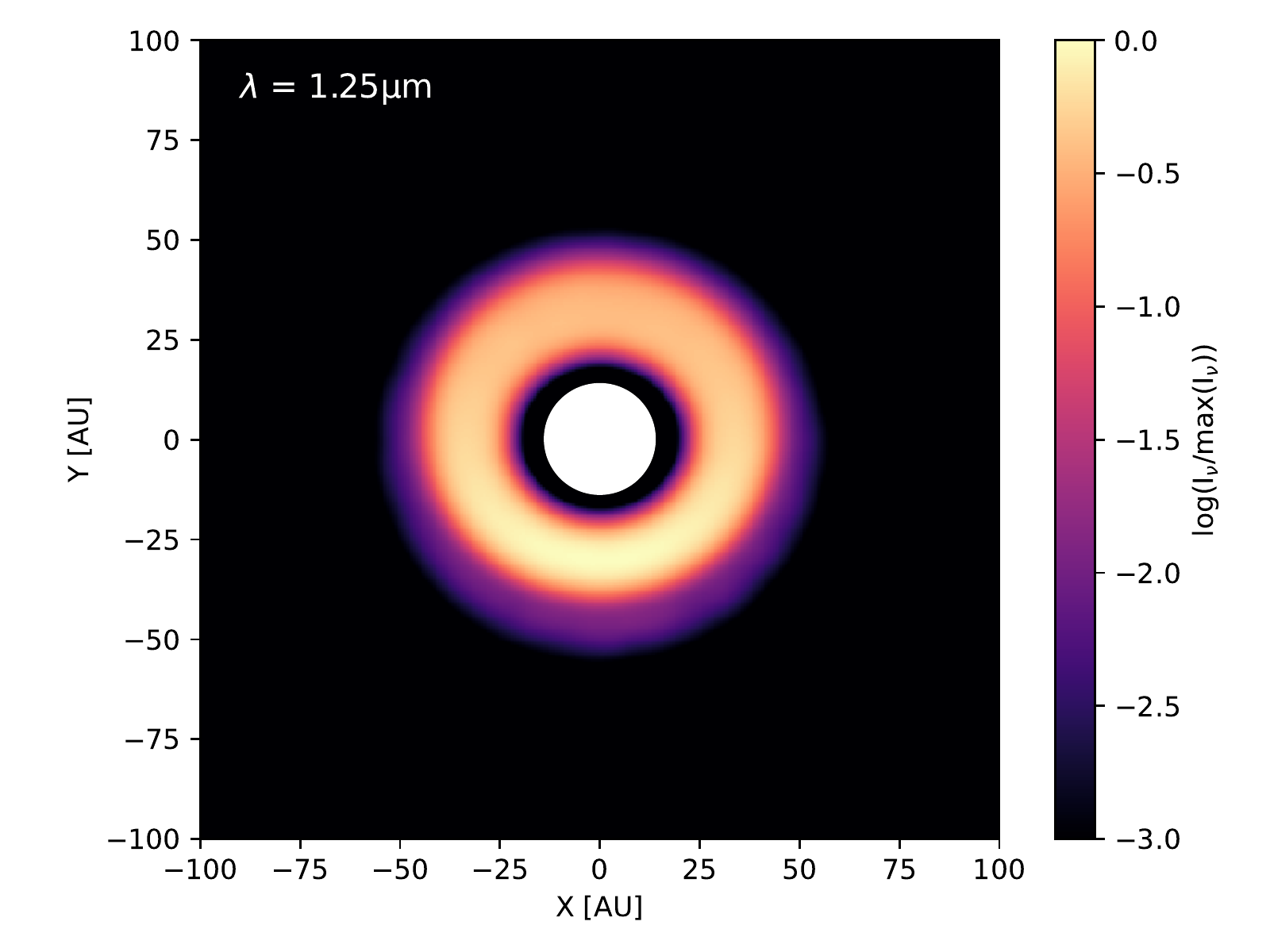}
 \caption{
 Synthetic images of the radiative transfer models for the dust size distribution presented in \autoref{sec_Results_DustDistribution} at $t = \SI{1.6}{Myr}$, both convolved with a Gaussian beam of $\SI{40}{mas}$.
 \textit{Top:} Continuum emission at $\lambda = \SI{1.3}{mm}$, the beam size is shown in the lower left corner.
 \textit{Bottom:} Scattered light brightness at $\lambda = \SI{1.25}{\mu m}$, normalized to the peak emission of the ring. The inner disk is covered by a \SI{0.1}{''} coronograph. 
 }
 \label{Fig_RadiativeTransfer}
\end{figure}
\begin{figure}
\centering
\includegraphics[width=90mm]{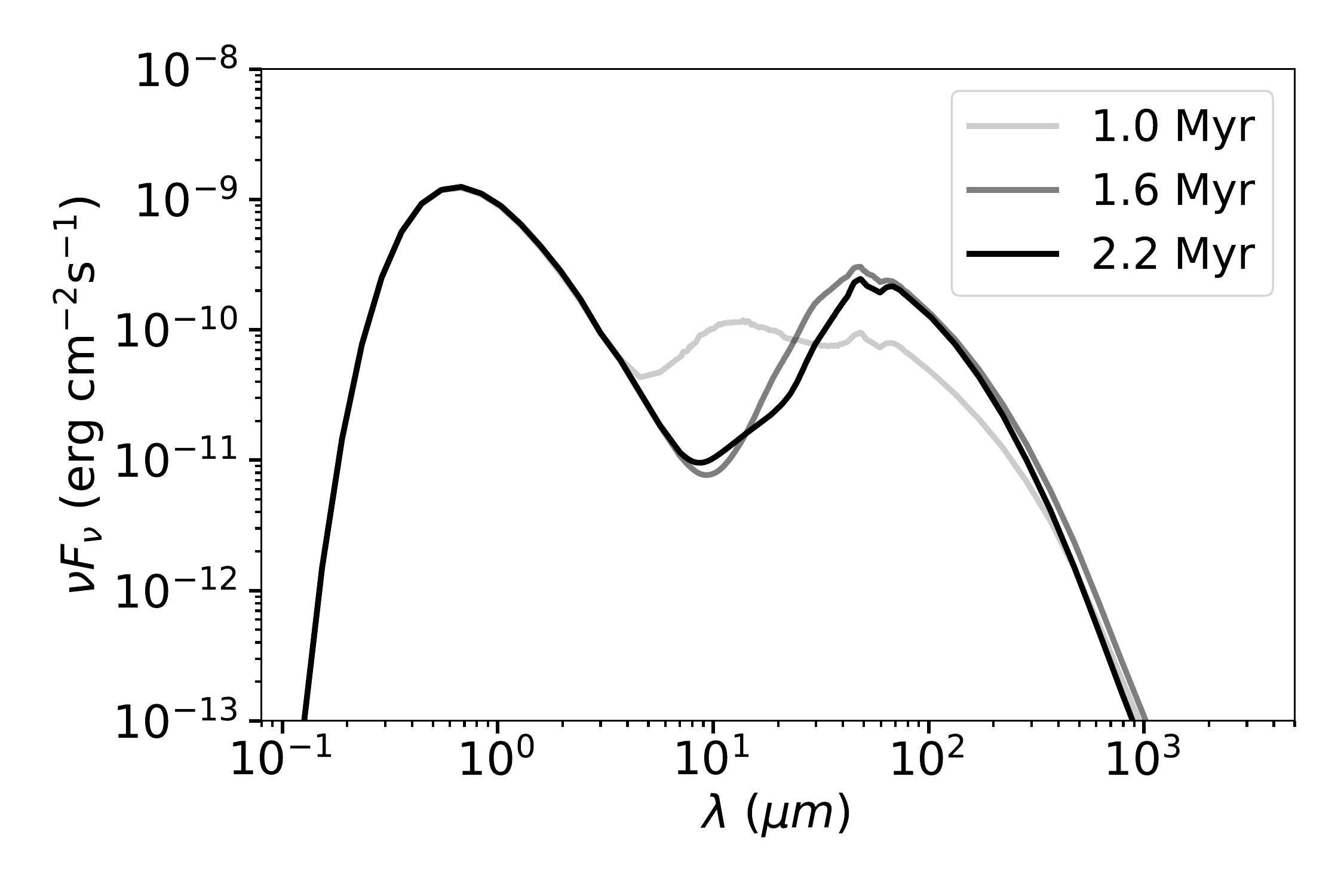}
 \caption{
 Spectral energy distribution for the disk model described in \autoref{sec_Results_DustDistribution} at three different times. The disk is assumed to be at a distance of $\SI{140}{pc}$ and at an inclination of $20^{\circ}$. 
 }
 \label{Fig_Spectra}
\end{figure}
Using the dust size distributions obtained with \texttt{DustPy} in \autoref{sec_Results_DustDistribution}, we perform radiative transfer calculations using the code \texttt{RADMC-3D}.\par
We calculate the opacities for each grain size using the program \texttt{OpTool}\footnote{  \href{https://github.com/cdominik/optool}{github.com/cdominik/optool}, version 1.7.2} assuming compact spheres and a dust composition consisting of water ice \citep{warren&brandt08}, astronomical silicates \citep{draine2006}, troilite and refractory organics \citep{henning1996} as stated in \cite{birnstiel18}. 
To account for the problems caused by the strong forward scattering peak in the phase function of large grains and the limited spatial resolution or our model, we set the scattering opacity to an upper limit within the first $5^\circ$ of the phase function. 
We assumed the solar type star to be a blackbody point source and used $10^7$ photons for our radiative transfer calculations.
First, the 3D dust temperature structure is computed which is then used to calculate the synthetic images and SEDs with the full treatment of anisotropic scattering with polarization \citep{Kataoka2015}, as shown in \autoref{Fig_RadiativeTransfer} and \autoref{Fig_Spectra}, respectively. The disk is assumed to have an inclination of 20° and is observed at a distance of 140 pc, which is the mean value of the nearest star forming regions. Finally, both images were convolved using a Gaussian point-spread-function with full-width at half-maximum (FWHM) of 0.04" $\times$ 0.04" to mimic the current angular resolution obtained by instruments like the \dbquote{Spectro-Polarimetric High-contrast Exoplanet Research}(SPHERE) at the Very Large Telescope (VLT), and the Atama Large Millimeter/sub-millimeter Array (ALMA).\par
\autoref{Fig_RadiativeTransfer} (top panel) shows the \SI{1.3}{mm} continuum emission of our model after the gap opening (at $\SI{1.6}{Myr}$). 
We observe a signal coming from the inner disk ($r \approx \SI{10}{AU}$) where the dead zone is located, and from a ring of dust located at $r \approx \SI{40}{AU}$, which corresponds to the pressure trap at the outer edge of the photoevaporative gap. 
The region in between is devoid of any emission, as expected from the dust distribution seen in \autoref{Fig_DustSizeDist}.\par
The total emission in the \SI{1.3}{mm} continuum disk is $F_\textrm{mm} = \SI{14.2}{mJy}$. 
In terms of the classification of transition disks according to their fluxes \citep[][]{Owen2012}, our model would be classified as a mm-faint disk ($F_\textrm{mm} \leq \SI{30}{mJy}$). 
This is at odds with the observed properties of disks with large gaps ($r_\textrm{gap} \gtrsim \SI{20}{AU}$) and high accretion rates ($\dot{M}_\textrm{g} \sim \SI{e-9}{M_\odot\, yr^{-1}}$), which are more likely to have mm fluxes above $\SI{30}{mJy}$ \citep[see][Figure 5]{Owen2016_review}. 
We discuss the implications of the low millimeter flux found in our model in \autoref{sec_Discussion_RelicDisks}.\par
The scattered light image (\autoref{Fig_RadiativeTransfer}, bottom panel) shows a wide ring, that extends approximately between \SI{25}{} and \SI{50}{AU}.
There is a signal coming from the inner disk in the scattered light, which is mainly occluded by the chronograph that we assume (\SI{0.1}{''} in size, which is typical for observations with SPHERE).\par
\autoref{Fig_Spectra} shows the evolution of the SED. 
The SED presents a decrease in the IR emission between \SI{1.0}{Myr} and \SI{1.6}{Myr}, approximately between $\SI{5}{\mu m}$ and $\SI{30}{\mu m}$, which coincides with the gap opening and the depletion of micron-sized particles in the inner disk (see \autoref{Fig_DustSizeDist}). This feature is characteristic of transition disks, and this object would be classified as such from its SED \citep[][]{Strom1989, Espaillat2014}.
\section{Discussion}\label{sec_Discussion}
\subsection{Comparison with the observed transition disk population} \label{sec_Discussion_RelicDisks}
The discrepancy between the predictions of theoretical models and the observed population of transition disks has been an open problem for the last decade.
Population synthesis models including photoevaporation predict that $88\%$ to $97\%$ of transition disks should be non-accreting disks with large gaps \citep[][]{Owen2011, Picogna2019}, also called \dbquote{relic disks}. Observational constrains on the other hand, say that the fraction of these  non-accreting relics should be only around $3\%$ \citep{Hardy2015}. In the literature, this discrepancy can also be found as the \dbquote{relic disk problem}.\par
Recent works from \cite{Ercolano2010, Ercolano2018, Wolfer2019} suggest that photoevaporation is more effective in disks that are depleted from carbon and oxygen, which results in higher mass loss rates and a faster dispersal of the outer disk. The outcome of this approach is a lower fraction of relics (up to $45\%$), which is much lower than standard photoevaporative models, and better suited to explain observations.\par
Meanwhile, models including dead zones and grain growth have also been able to reproduce some of the features observed in transition disks, including the decrease in the NIR emission due to inefficient fragmentation, and ring like structures \citep[][]{Birnstiel2012_b, Pinilla2016}, however these models also fail to clear extended gaps in the millimeter continuum, because the drift timescales of the large grains become too long, or because the particle trapping occurs beyond the dead zone outer edge.\par
In this work we present a different approach, which consist on combining the photoevaporative dispersal by X-ray radiation, with a dead zone in the inner disk. This way, both processes complement each other by covering their respective weak spots: the dead zone takes care of extending the lifetime of the inner disk, keeping the accretion rates high, while photoevaporation carves extended gaps in the gas and dust components \citep[see also][]{Morishima2012}.\par
The strongest point of this model is that it produces a high fraction disks with large gaps ($r_\textrm{gap}\gtrsim \SI{20}{AU}$) and  high accretion rates ($\dot{M}_\textrm{g} \sim \SI{e-9}{M_\odot \,yr^{-1}}$) at the same time, predicting that up to $63\%$ of all transition disks should be accreting. 
This fraction is higher than previous population synthesis estimates, though it is still not high enough to match the observational constrains by itself \citep[][]{Hardy2015}, since we still find that $37\%$ of transition disks should be relics (i.e. non-accreting).\par
The predicted fraction of relic disks could be even lower if the effects of carbon depletion are considered, with the respective increase in the mass loss rate and faster dispersal of the outer disk \citep{Ercolano2018}, or if the disks inner regions are able to survive for longer times, by reaching a quasi-steady state during earlier stages of disk evolution for example (see \autoref{sec_Appendix_GasDust}).\par
Alternatively, it could also be that many disks become too faint to be observed if dust is removed efficiently, either by drift during early stages of disk evolution, or by a strong coupling with the photoevaporative wind \citep[][]{Owen2019}. 
While in our simulations the dust loss rate is always too low to affect the total dust mass, a more detailed modelling of the gas drag at the edge of the photoevaporative gap could yield higher dust loss rates at lower scale heights, after the dispersal of the inner disk.\par
In the context of the whole disk population (i.e. counting primordial and transition disks), our results seem to match the fraction of transition disks shown in \cite{Currie2011}, especially around $\SI{2}{Myrs}$, with the dead zone having little to no influence on whether a gap is opened or not. 
At later times, we do predict more transition disks than those observed, but this discrepancy could also be bridged if we consider that some of our disk should become fainter over time, again due to dust evolution, entrainment and drifting.\par
We should keep in mind that the accuracy of our predictions depends, of course, on the validity of our dead zone model (\autoref{eq_alpha_profile}), since the fraction of accreting transition disks varies greatly across different dead zone parameters (see \autoref{Table_DiskFraction}). We discuss this point in \autoref{sec_Discussion_DeadZone}, where we calculate what a self-consistent dead zone model would look like.\par

One apparent limitation of our model is that it is unable to produce enough transition disks with narrow gaps ($r_\textrm{gap} \lesssim \SI{10}{AU}$, see \autoref{Fig_ProbabilityDist_Param}), since by construction the combined effects of dead zones and photoevaporation favor opening a gap beyond the dead zone outer edge \citep{Morishima2012}. 
This seems at odds with the observations, which indicate there are approximately as many transition disks with gaps narrower than \SI{10}{AU}, as disks with wider gaps, with sizes between $\SI{10}{AU}$ and $\SI{100}{AU}$ \citep[][]{vanderMarel2016}.
However, most of the gaps in transition disks with sizes smaller than \SI{10}{AU} are still unresolved by observations, and they are characterized by SED fitting. This means that the current gap sizes reflect the deficit of small dust grains in the inner regions, but may not correspond to an actual gap in the gas or in the millimeter continuum.
We could still predict this type of narrow gaps within our framework, if we consider that dust growth in dead zones can deplete the inner regions of small grains, and create SED signatures that are alike to that of a transition disk \citep[][]{Birnstiel2012_b}.
Therefore, a disk could display a narrow gap signature in their SED due to a dead zone at earlier times, and a wide gap signature in both their SED and millimeter continuum at later stages due to the combined dead zone and photoevaporation effects described in \cite{Morishima2012} and this work.\par
Besides the accretion rate and gap sizes, the observed transition disks seem to be divided in two groups according to their millimeter flux $F_\textrm{mm}$ \citep{Owen2012}. Disks with smaller gaps tend to be \dbquote{mm-faint} ($F_\textrm{mm} < \SI{30}{mJy}$), while disks with larger cavities are \dbquote{mm-bright} ($F_\textrm{mm} > \SI{30}{mJy}$), with the fluxes re-scaled to a distance of \SI{140}{pc} \citep[see][Figures 3 and 5]{Owen2016_review}.
Instead, the millimeter flux of our dust evolution model is only $F_\textrm{mm} = \SI{14.2}{mJy}$, despite presenting a large cavity and a high accretion rate. Simulations with different parameters that were not included in this paper also show similar fluxes, around $\SI{10}{mJy}$. 
This suggest that the dust content in our disks is relatively low, in comparison with with the observed transition disks, or that we are underestimating the dust opacity.\par
We can explain why our disks appear to be dust depleted by looking at our initial gas density profile (\autoref{eq_LBPprofile}), which is smooth and does not have any substructure. 
In such a disk the dust particles can drift inwards and into the star very quickly, since there are no pressure maxima to act as local dust traps \citep[][]{Pinilla2012_b}.
In contrast, observations of protoplanetary disks show that these are rich in substructures such as rings and spirals \citep[e.g.][]{ALMA2015, Andrews2018}, indicating that the underlying gas profile is not smooth.
Recent models also indicate that dust traps are necessary in order to explain the size luminosity relationship observed in protoplanetary disks \citep[][Zormpas et al. in prep.]{Tripathi2017}. 
Including pressure bumps in our model at early times, such as a gap created by a planet, would help our disk to retain a higher fraction of dust, increasing its millimeter emission and reducing the discrepancy with observations \citep[][]{Pinilla2012, Pinilla2020}.\par
Despite this limitation in our model, it seems that the presence of a dead zone in the inner regions is a promising ingredient to explain the observed transition disk population.
For a follow up work, we intend to quantify the millimeter continuum emission of disks undergoing photovaporative dispersal, and test if it is possible to retain enough solids to match the observed fluxes from transition disks through trapping \citep[][]{Pinilla2020}.\par
Additional improvements for the model would be to include stars of different mass, considering their respective photoevaporation mass loss rates \citep{Komaki2021}, the evolution of the stellar luminosity during the first Myr after formation \citep[][]{Kunitomo2021}, and the effect of carbon depletion, which leads to stronger photoevaporation rates \citep[][]{Ercolano2018, Wolfer2019}.
%
%
%
\subsection{Long-lived inner disk and comparison with observations} \label{sec_Discussion_InnerDisk}
The large diversity in the SED morphology of transition disks \citep[e.g.,][]{cieza2007} provides evidence that the cavity of some of these disks is not completely depleted in dust. 
Some of the SEDs of transition disks exhibit a strong excess in the NIR wavelength range \citep[e.g.,][]{Espaillat2010, vanderMarel2016}, implying a compact optically thick inner disk very close to the star.\par
Observational efforts have been made to directly observe and characterize the inner disk of transition disks. For example, interferometric observations at the NIR have spatially resolved a compact inner disk in some transition disks \citep[e.g.][]{Benisty2010, Olofsson2013, Perraut2019_Gravity, Bouarour2020_Gravity}. Recent observations with ALMA also detect and in few cases resolve the inner disk of some transition disks, meaning that these inner disks are probably rich in millimeter-sized particles, or that there is an equivalent amount of smaller grains that lead to a similar signal in the millimeter continuum \citep[e.g.][]{hendler2018, Pinilla2019, keppler2019, Francis2020, Rosotti2020}.\par
The physical mechanism responsible for the observed dust in the inner disk of transition disks is an open question, and it is unclear how it can survive on timescales of millions of years. For the formation of the large gaps in these objects, mainly three possibilities are debated in the literature: embedded planets, photoevaporation, and dead zones. However, each of these scenarios have their own challenges to explain a long-lived inner disk as observed at different wavelengths. \par
On one hand, the planet-disk interaction models may allow to have partial dust filtration at the outer edge of the gap \citep[][]{Pinilla2012, Weber2018}, such that the inner disk can be replenished with small dust (micron-sized) from the outer disk. However, in this scenario these small particles grow very efficiently in the inner disk once they cross the gap and thus they are lost towards the star due to efficient drift. This problem can be mitigated by including the effect of the snow line, that can change the velocities for which particles fragment \citep[][]{Wada2011, Gundlach2011, Gundlach2015}, and hence slow down their radial drift \citep{Birnstiel2010, Pinilla2016b, Drazkowska2017}. These models of planets with the snow line are capable of reproducing the NIR excess seen in the SEDs of some transition disks, but they fail in reproducing a detectable inner disk at millimeter wavelengths \citep[see Figure A.1 in][]{Pinilla2016b}.\par
On the other hand, models of dead zones acting alone can reproduce an inner dusty disk, but it is expected to be much fainter than the ring-like emission in the outer disk at millimeter wavelengths \citep[see Fig. 7 and Fig. 8 in][]{Pinilla2016}. 
Additionally, in these models there is not a clear depletion of the gas surface density inside the mm-cavity, which seems to be highly depleted from observations of $^{12}$CO and its isotopologues \citep[][]{marel2016b}. 
To solve this problem, it was suggested that the inclusion of an MHD wind in the simulations could reduce the gas surface density in the inner regions, but would worsen the problem of keeping a long-lived inner disk.\par
Finally, photoevaporation models previous to this work predicted highly depleted cavities, in both gas and dust \citep[][]{Alexander2007, Owen2019}, but these ignored the combined effect of a potential dead zone in the photoevaporative dispersal process.\par
The models presented in this paper can explain a dusty inner disk that remains millions of years, and that is bright at millimeter emission.\par 
The maximum grain size in the inner disk is limited by drift, and therefore by the local dust-to-gas ratio, which is gradually decreasing with time. As a consequence the radial drift velocity of the particles in the inner disk also decreases, prolonging their lifetime (\autoref{Fig_DustSizeDist}).
This solves the conundrum of the origin and survival of these inner disks in transition disks.\par
From our synthetic observations at millimeter wavelength (\autoref{Fig_RadiativeTransfer}), we find that the surface brightness in the inner disk is slightly fainter (by a factor of a few) than the ring at the outer edge of the photoevaporative gap. 
A similar feature has been observed  in some transition disks \citep[e.g., TCha and CIDA1,][respectively]{hendler2018, Pinilla2021}, where the inner disk has a comparable brightness to the outer disk.
The millimeter sized particles in the inner regions extend until the outer edge of the inner disk, as given by the gas distribution, indicating that the dust can be retained for long timescales (along with the gas component) without the need of additional dust traps.
%
%
\subsection{Validity of the dead zone model} \label{sec_Discussion_DeadZone}
\begin{figure}
\centering

\includegraphics[width=90mm]{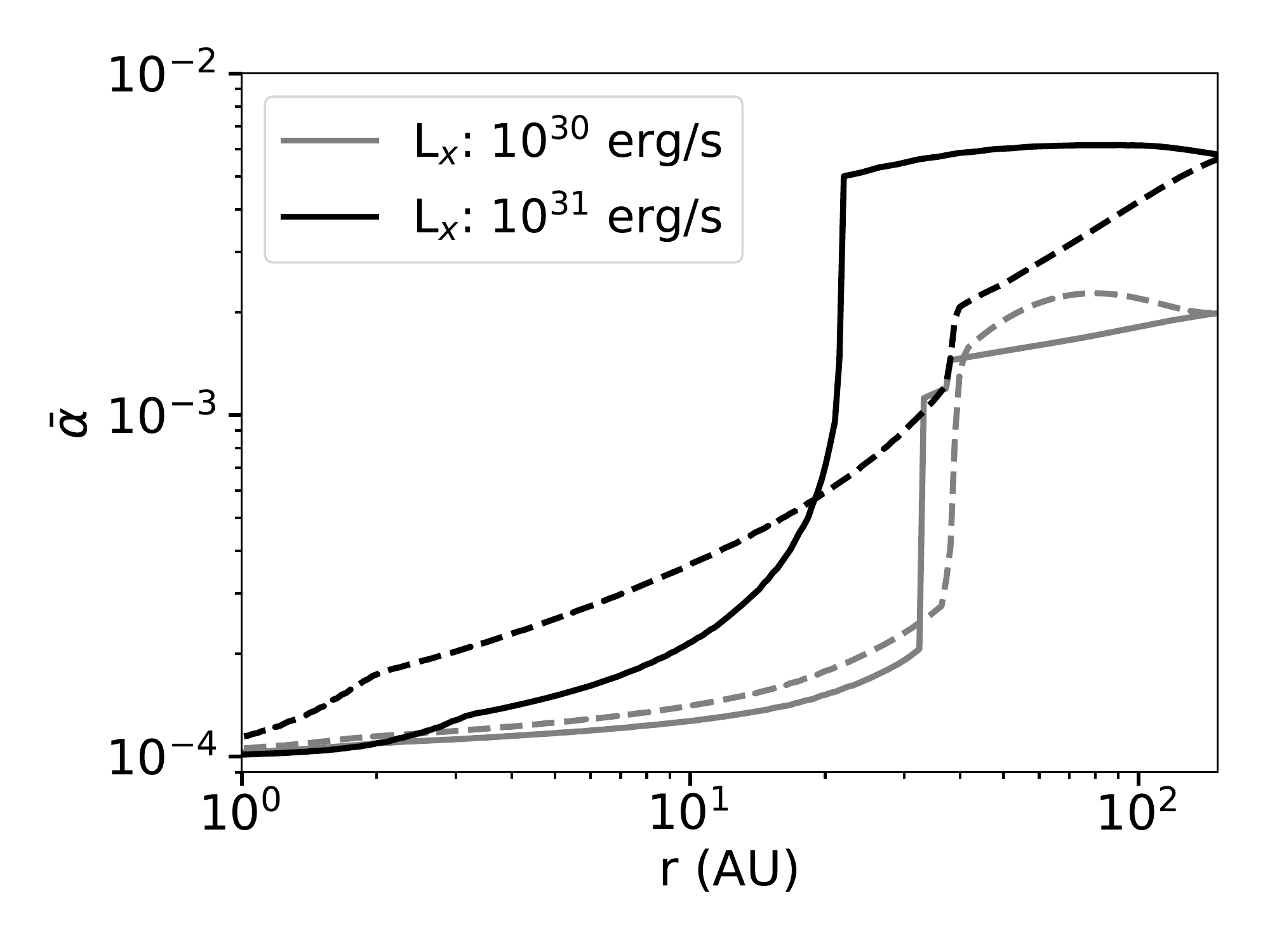}
 \caption{
 Effective turbulence $\bar{\alpha}$ profile for different X-ray luminosities, derived using the MRI model described in \cite{Delage2021}. The solid and dashed lines correspond to the steady-state scenario and \cite{Lynden-Bell1974} scenario, respectively. See \autoref{sec_Discussion_DeadZone}.
 }
 \label{Fig_ValidityAlpha}
\end{figure}
In this work, we have modelled the presence of a static dead zone by using the $\alpha(r)$ profile described in \autoref{eq_alpha_profile}, which is defined by the free parameters $\alpha_\textrm{dz}$, $\alpha_\textrm{a}$, and $r_\textrm{dz}$.\par 
Our parameter space study in \autoref{sec_Results_Gas} shows that the fraction of accreting transition disks is heavily influenced by the turbulence in the dead zone $\alpha_\textrm{dz}$, and the dead zone extent $r_\textrm{dz}$.
For this reason we are interested to see if a more complex model could produce similar profiles in terms of the dead zone turbulence and radial extent.\par
We perform a quick validation test using the framework developed in \cite{Delage2021}. 
The authors have built a 1+1D magnetically-driven disk accretion model for the outer disk regions ($r \gtrsim 1\,$AU), where the local mass and angular momentum transport are assumed to be controlled solely by the MRI and hydrodynamic instabilities such as e.g. the VSI. From such a model, we can derive an effective turbulent parameter $\bar{\alpha}$ that is self-consistently determined from detailed considerations of the MRI and the non-ideal MHD effects (Ohmic resistivity and Ambipolar diffusion) -given stellar properties (mass and luminosity), disk mass and dust properties. Their model accounts for the main physical processes happening in the outer regions of protoplanetary disks: (1) irradiation from the forming star; (2) dust settling; (3) ionization from stellar X-rays, galactic cosmic rays and the decay of short/long-lived radionuclides; (4) disk chemistry.
Using their framework, we investigate two scenarios. First, the effective turbulence $\bar{\alpha}$ is calculated from a gas surface density following the standard self-similar solution from \cite{Lynden-Bell1974}. Since this scenario leads to a disk structure out of equilibrium (the accretion rate is radially non-constant, implying that some regions of the disk can accrete faster than others), we consider a second scenario where the effective turbulence $\bar{\alpha}$ is derived by assuming steady-state accretion. In the latter, the gas surface density is self-consistently derived through an iterative process, alongside the effective turbulence $\bar{\alpha}$, to ensure the accretion rate to be radially constant. For each scenario, we consider two X-ray luminosities ($\SI{e30}{}$ and $\SI{e31}{erg\, s^{-1}}$). The stellar and disk mass are the same as described in \autoref{sec_SetupDisk}. We further assume a constant dust-to-gas ratio of $\epsilon = 0.01$ as well as grains of size $a = \SI{1}{\mu m}$, consistent with the setup described in \autoref{sec_SetupDust}.\par
\autoref{Fig_ValidityAlpha} shows the $\alpha(r)$ profiles derived for the four simulations, using the framework described in \cite{Delage2021}. Depending on the X-ray luminosity used or the scenario considered, the salient results are as follows: the dead zone extent $r_\textrm{dz}$ ranges from $\SI{22}{AU}$ to $\SI{40}{AU}$, where the turbulence quickly changes from a low to high regime; the mean turbulence in the dead zone $\alpha_\textrm{dz}$ ranges from $\SI{1.3e-4}{}$ to $\SI{3.7e-4}{}$; and the mean turbulence in the MRI-layer $\alpha_\textrm{a}$ ranges from $\SI{1.8e-3}{}$ to $\SI{5.7e-3}{}$.\par
Based on these calculations, we find that the dead zone radial extent should be larger than the one used in our models, which in principle would mean that the fraction of accreting transition disks could be better represented by populations with $r_\textrm{dz} = \SI{20}{AU}$.
We also find that the mean values for the turbulence in the dead zone are covered by our parameter space exploration. Furthermore, the highest value found ($\alpha_\textrm{dz} = \SI{3.7e-4}{}$) would suggest that our choice of $\alpha_\textrm{dz} = \SI{5e-4}{}$ is quite marginal, and for moderate X-ray luminosities ($L_x = \SI{e30}{erg\, s^{-1}}$) the dead zone turbulence remains between $\alpha_\textrm{dz} = \SI{e-4}{}$ and $\SI{2e-4}{}$.\par
We expect that a population synthesis model using the dead zone model from \citep{Delage2021} would also present gaps that extend beyond 20 - 40 AU, and a fraction of accreting disks above $12\%$ at least, and possibly closer to $55\%$
(see the models with $\alpha_\textrm{dz} = \SI{1e-4}{}$ and $\SI{3e-4}{}$, for $r_\textrm{dz} = \SI{20}{AU}$, \autoref{Table_DiskFraction}).\par
Finally, we note that our choice for $\alpha_\textrm{a}$ underestimates the mean values found by the more complex model. A higher turbulence in the MRI-active layer could result in the photoevaporative gap to open earlier, which we expect to result in a higher dust content in the outer disk, without decreasing the fraction of accreting disks found in this work.
Indeed, $\alpha_\textrm{dz}$ and $r_\textrm{dz}$ are the key parameters here, since these determine the lifetime and viscous timescale of the inner disk.
%
%
\section{Summary}\label{sec_Summary}
In this work we present a disk evolution model that includes X-ray photoevaporative dispersal and a dead zone prescription, in order to explain the accretion rates and extended gap sizes observed in transition disks.\par
Our population synthesis study shows that considering a dead zone in the inner regions can easily result in transition disks with extended gaps ($r_\textrm{gap} \gtrsim \SI{20}{AU}$), and long lived inner disks capable of sustaining high accretion rates ($\dot{M}_\textrm{g} \sim \SI{e-9}{M_\odot\, yr^{-1}}$) during the dispersal process.\par
This is a result of the differential evolution of the inner and outer disk. Because the outer turbulent disk evolves on shorter viscous timescales, its accretion rate decreases faster and enters into the photoevaporative dispersal regime earlier. Meanwhile, the viscous timescale of the inner disk is longer due to the low turbulence in the dead zone, allowing it to sustain a relatively high accretion rate.\par
For dead zone properties of $\alpha_\textrm{dz} = \SI{e-4}{}$ and $r_\textrm{dz} = \SI{10}{AU}$, we predict that $63\%$ of transition disks should still be accreting, with $\dot{M}_\textrm{g} \gtrsim \SI{e-11}{M_\odot\, yr^{-1}}$. We also find that half of these accreting transition disks display high accretion rates of $\dot{M}_\textrm{g} \gtrsim \SI{5.e-10}{M_\odot\, yr^{-1}}$.
This means that photoevaporative disk dispersal could, in fact, explain the observed accreting transition disks with large gaps, though it is still necessary to explain why we do not detect the predicted fraction of non-accreting disks.\par
While our model does not explicitly predict transition disks with narrow gaps ($r_\textrm{dz} \lesssim \SI{10}{AU}$), we suggest that these could still be explained through other processes, such as dust growth within a dead zone during the earlier stages of disk evolution, which would then evolve into wide photoevaporative gaps at later stages.\par
From our dust evolution simulations, we learn that the inner disk retains a dust component consisting of large millimeter to centimeter sized grains, while the dust in the outer disk forms a ring at the edge of the photoevaporative gap, with grains sizes between a micrometer and a few millimeters.
Radiative transfer calculations based on our dust distribution model show
an inner and outer disk in the $\SI{1.3}{mm}$ continuum, with the inner disk being fainter by a factor of a few, and that the SED shows an emission deficit at near and mid-infrarred wavelengths, which is a characteristic feature of transition disk spectra.\par
While the millimeter flux in our model appears to be fainter than the ones observed in transition disks with large cavities, this could be solved by including dust traps in the outer disk (such as the ones caused by planets), or more vigorous photoevaporation rates (as those predicted for carbon depleted disks), in order to retain a higher fraction of solids during the early stages of disk evolution.\par
In future studies we plan to focus on the observability of photoevaporating disks by extending our analysis of the dust component, to determine if the predicted non-accreting disks could be too faint to be observed, and if additional substructures could reproduce the observed population of millimeter bright transition disks with high accretion rates.\par
In summary, our model including dead zones and photoevaporation is a good candidate to explain several features observed in transition disks, from the high accretion rates, to the large gaps, and even the inner disks observed in the millimeter continuum.

\begin{acknowledgements}
We would like to thank the anonymous referee for the useful feedback, that greatly improved the quality of this work.
The authors acknowledge funding from the Alexander von Humboldt Foundation in the framework of the Sofja Kovalevskaja Award endowed by the Federal Ministry of Education and Research, from the European Research Council (ERC) under the European Union’s Horizon 2020 research and innovation programme under grant agreement No 714769, and by the Deutsche Forschungsgemeinschaft (DFG, German Research Foundation) under Germany's Excellence Strategy – EXC-2094 – 390783311 and Ref no. FOR 2634/1.
\end{acknowledgements}
\bibpunct{(}{)}{;}{a}{}{,} 
\bibliographystyle{aa} 
\bibliography{aa_main_mgarate.bbl} 

%
%
%
\begin{appendix}
\section{Models in quasi-steady state} \label{sec_Appendix_GasDust}
\begin{figure*}
\centering
\includegraphics[trim={160px 75px 270px 135px},clip,width=0.99\textwidth]{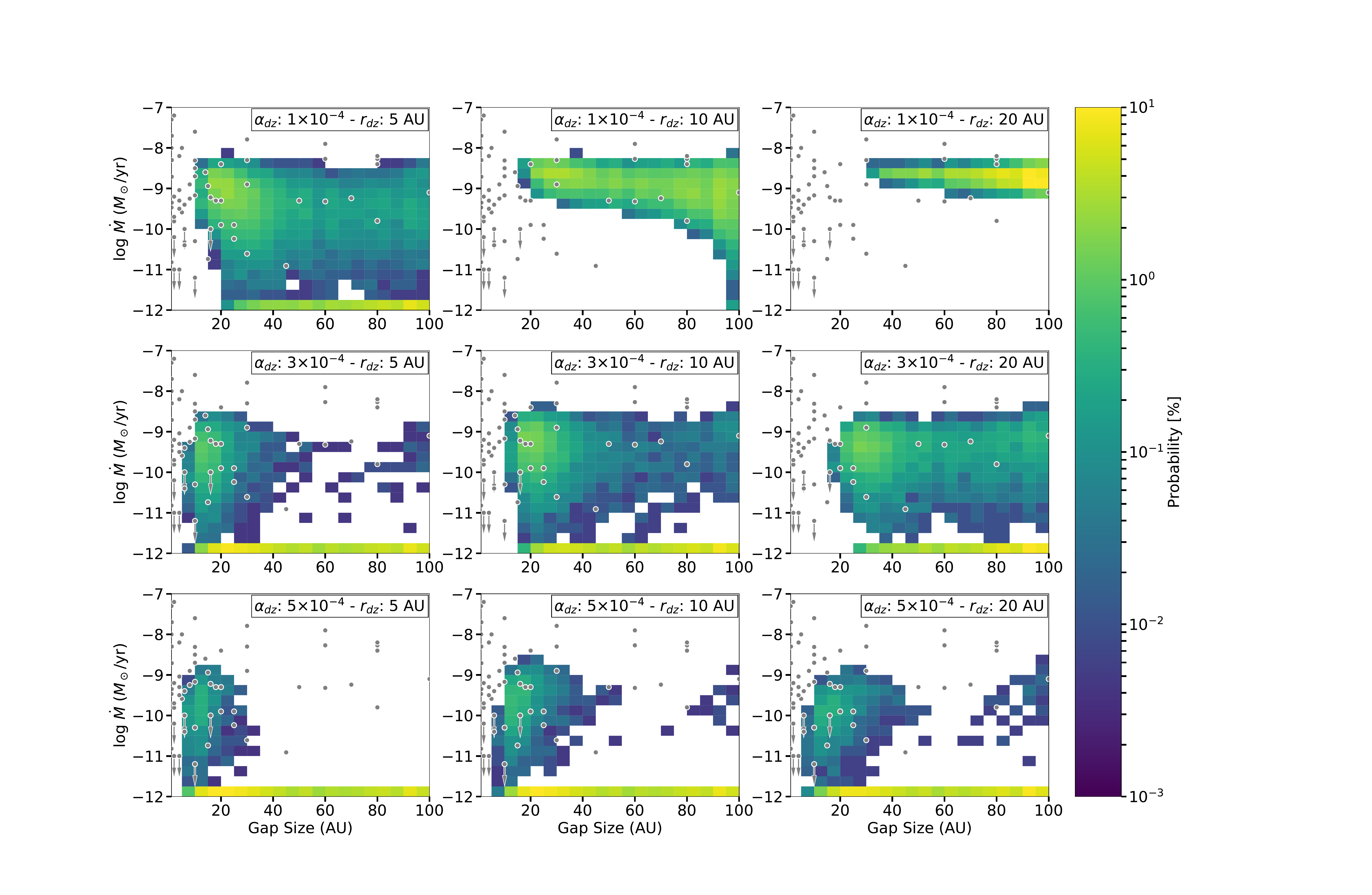}
 \caption{
 Same as \autoref{Fig_ProbabilityDist_Param}, using the initial condition described in \autoref{eq_LBPprofile_Scaled}.
 }
 \label{Fig_ProbabilityDist_ParamSteadyState}
\end{figure*}


\begin{table}
 \caption{Percentage of transition disks with $\dot{M}_g > \SI{e-11}{M_\odot\, yr^{-1}}$ for different dead zone models, using the initial condition in \autoref{eq_LBPprofile_Scaled}.}
 \label{Table_DiskFraction_SteadyState}
 \centering
  \begin{tabular}{ c | c | c}
    \hline \hline
    \noalign{\smallskip}
    $\alpha_\textrm{dz}$ & $r_\textrm{dz}$ (AU) & Accreting Disk Fraction (\%)\\
    \hline
    \noalign{\smallskip}
     & \SI{20}{} & 100.0 \\
    \SI{e-4}{} & \SI{10}{} & 99.7 \\
     & \SI{5}{} & 52.8  \\
    \hline
    \noalign{\smallskip}
     & \SI{20}{} & 35.1 \\
    \SI{3e-4}{} & \SI{10}{} & 25.0 \\
     & \SI{5}{} & 10.9  \\
    \hline
    \noalign{\smallskip}
     & \SI{20}{} & 4.7 \\
    \SI{5e-4}{} & \SI{10}{} & 6.1 \\
     & \SI{5}{} & 3.2  \\
    \hline
  \end{tabular}
\end{table}
\begin{figure}
\centering
\includegraphics[width=95mm]{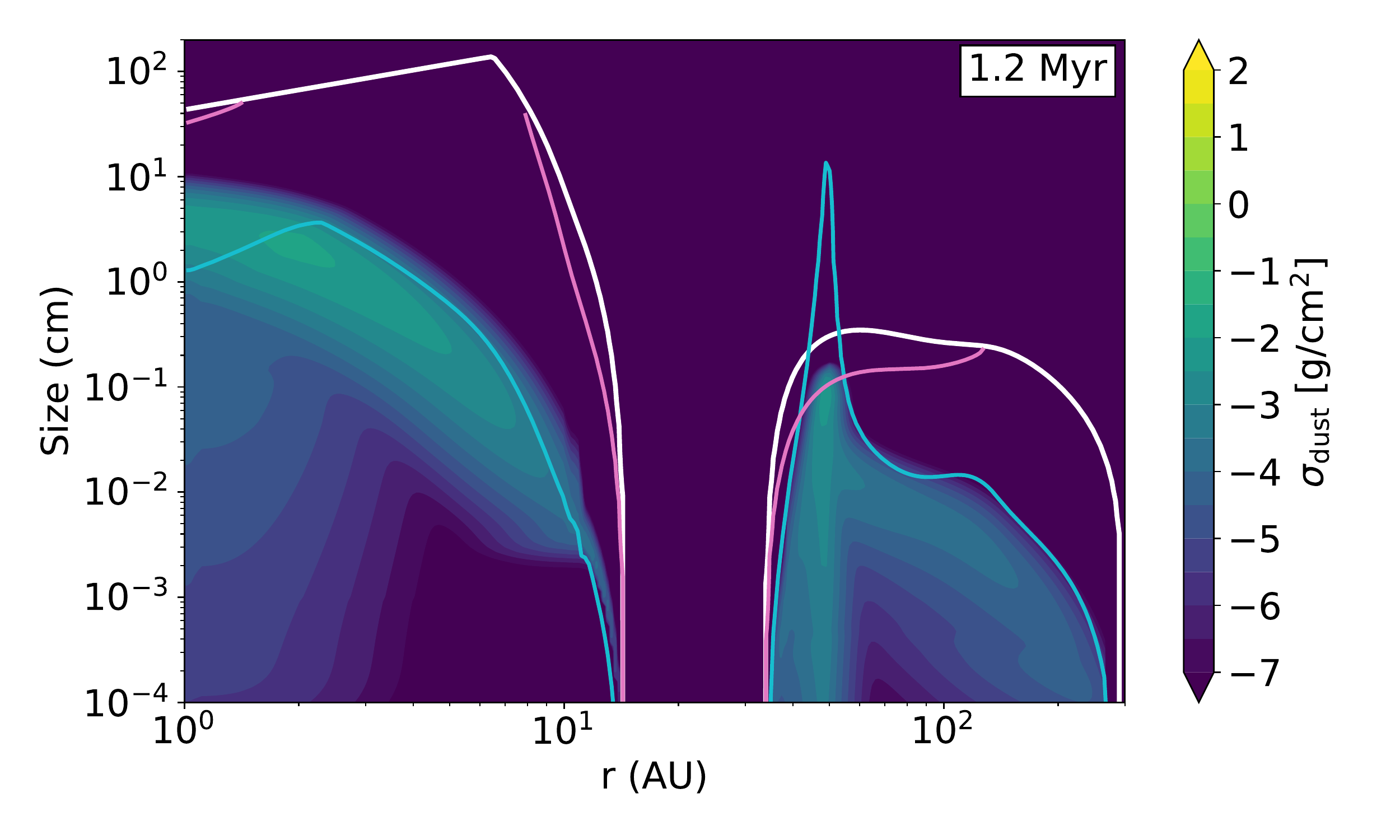}
 \caption{
 Dust size distribution at $\SI{1.2}{Myr}$ for a disk evolution model initialized in quasi-steady state (\autoref{eq_LBPprofile_Scaled}). All the other parameters are the same as in \autoref{Fig_DustSizeDist}.
 }
 \label{Fig_DustDistribution_SteadyState}
\end{figure}
\begin{figure}
\centering
\includegraphics[width=95mm]{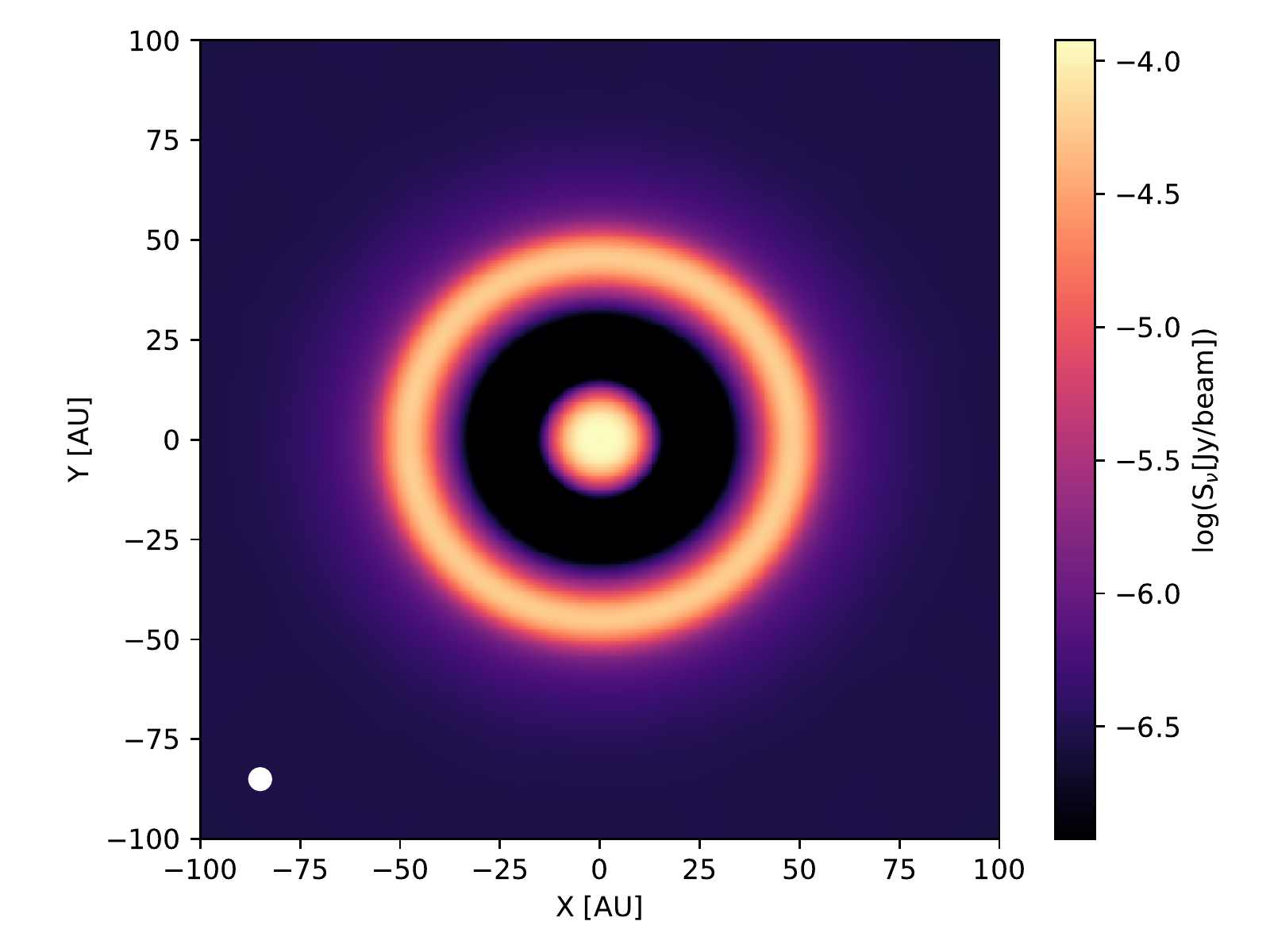}
 \caption{
 Radiative transfer model of the continuum emission at $\lambda = \SI{1.3}{mm}$, for the dust size distribution presented in \autoref{Fig_DustDistribution_SteadyState}, and convolved with a Gaussian beam of $\SI{40}{mas}$. The beam size is shown in the lower left corner.
 }
 \label{Fig_RadiativeTransfer_SteadyState}
\end{figure}
In this appendix we show an additional set of population synthesis models with a different initial condition, to complement the results of section \autoref{sec_Results_Gas}.\par
For this work, we studied disks that are initialized with a \cite{Lynden-Bell1974} profile, which consist on a power law with an exponential decay. When the turbulence profile is constant, these disks evolve in a so called \dbquote{quasi-steady state}, which is characterized by a constant accretion rate in the inner regions, that decays uniformly over the viscous timescale. 
However, if a disk has a non-uniform turbulence profile, such as the one created by a dead zone (see \autoref{Fig_AlphaModel}), this results in a variable accretion profile, that is not in local equilibrium (see the initial condition of \autoref{Fig_AccretionEvolution}, top panel).\par
In order to test the effect of the initial condition, we repeat our population synthesis study as described in \autoref{sec_SetupPopulation}, but now assuming the material is initially distributed such that the inner regions evolve in quasi-steady state, using a scaled version of the \cite{Lynden-Bell1974} profile:
\begin{equation} \label{eq_LBPprofile_Scaled}
    \Sigma_\textrm{g}(r) = \Sigma_0 \left(\frac{r}{r_c}\right)^{-1} \exp(-r/r_c)  \frac{\alpha_\textrm{a}}{\alpha(r)}.
\end{equation}
The scaling factor $\alpha_\textrm{a}/\alpha(r)$ results in a profile that has a higher density in the dead zone region, and a lower density in the outer regions, when compared to the standard \cite{Lynden-Bell1974} solution. The normalization factor $\Sigma_0$ is again defined such that the integrated mass is $M_\textrm{disk} = \SI{1}{M_\odot}$.\par
\autoref{Fig_ProbabilityDist_ParamSteadyState} shows the corresponding distribution of the gas accretion rate and gap sizes of transition disks, and \autoref{Table_DiskFraction_SteadyState} shows the fraction accreting transition disks using the initial condition from \autoref{eq_LBPprofile_Scaled}.
From these new results we immediately notice an increase in the abundance of transition disks accreting at high accretion rates. 
In particular, we find that for the model with $\alpha_\textrm{dz} = \SI{e-4}{}$ and $r_\textrm{dz} = \SI{10}{AU}$, the predicted fraction of accreting transition disks jumped from $62.9\%$ to $99.7\%$, with approximately half of those disks displaying accretion rates above $\SI{e-9}{M_\odot yr^{-1}}$.
Another difference is that now, using the scaled initial condition, the populations with more extended dead zones ($r_\textrm{dz} = \SI{20}{AU}$) tend to display the highest fraction of accreting transition disks (for $\alpha_\textrm{dz} \leq \SI{3e-4}{}$).\par
Both the overall increment in the fraction of accreting transition disks, and its systematic increase with the dead zone size $r_\textrm{dz}$, can be explained if we think that the new initial condition represents disk in which the inner regions are \dbquote{saturated} of material, that is, they have the required amount of gas to be in local steady state. 
In this scenario, the inner disk has the maximum possible mass by the time photoevaporation opens a gap, and therefore it can survive for the longest possible time.\par
Of course, the question now would be if the quasi-steady state initial condition described by \autoref{eq_LBPprofile_Scaled} can be met. 
From our results in \autoref{Fig_SurfaceDensityEvolution}, we see that in the presence of a dead zone, a disk initialized with a \cite{Lynden-Bell1974} profile slowly evolves into a quasi-steady state profile, like the one described by \autoref{eq_LBPprofile_Scaled} (though the exact shape might vary), but with a lower normalization factor $\Sigma_0$  and larger characteristic radius $r_c$, since some material must have been accreted, and the outer regions have expanded due to the transport of angular momentum.\par
We expect that disks with short viscous timescales in their inner regions would reach the quasi-steady state solution more easily. 
Accordingly, we find that the disk populations with smaller dead zones and higher turbulence parameters are be less affected by the choice of initial condition (see \autoref{Table_DiskFraction} and \autoref{Table_DiskFraction_SteadyState}), since they have enough time to reach the quasi-steady state regime before the photoevaporation opens a gap and disconnects the inner and outer disk from each other. 
In contrast, disks with extended dead zones are more affected by the choice of initial condition, since their viscous timescales are longer (of several million years).\par
Other factor that should affect whether a disk reaches the quasi-state state solution is the disk mass, since a more massive disk would be able to transport more material into the inner regions before photoevaporation becomes relevant for its global evolution, though the corresponding normalization factor would be lower, depending on how much material was accreted into the star.\par
Finally, one reason why it might be preferable to draw our conclusions from the standard \cite{Lynden-Bell1974} profile, instead from the quasi-steady state profile in \autoref{eq_LBPprofile_Scaled}, is that the dead zones could be subject to thermal and gravitational instabilities during the early stages of disk evolution \citep[][]{Martin2011}, where the turbulence is cyclically reactivated in the inner regions, resulting in accretion outbursts, and for the accumulated material to quickly flushed into the star \citep[][]{Kretke2009, Audard2014, Garate2019}. With this point in mind, we can think of the results based on the quasi-steady state initial condition as an upper limit to the fraction of accreting transition disks.\par
To conclude, we also show the effect of the quasi-steady state initial condition on the dust distribution and millimeter continuum synthetic observations in \autoref{Fig_DustDistribution_SteadyState} and \autoref{Fig_RadiativeTransfer_SteadyState}, also using an X-ray luminosity of $L_x = \SI{e31}{erg s^{-1}}$, $\alpha_\textrm{dz} = \SI{e-4}{}$, and $r_\textrm{dz} = \SI{10}{AU}$, but taking the snapshot at $\SI{1.2}{Myr}$, since the gap opens earlier when using the initial condition from \autoref{eq_LBPprofile_Scaled}. The main difference is that now the surface brightness of the inner disk is comparable to that of the outer ring in the millimeter continuum, as observed in some transition disks \citep[][]{hendler2018, Pinilla2021}.

\end{appendix}

\end{document}